\catcode`\@=11
{\count255=\time\divide\count255 by 60
\xdef\hourmin{\number\count255}
\multiply\count255 by-60\advance\count255 by\time
\xdef\hourmin{\hourmin:\ifnum\count255<10 0\fi\the\count255}}
\def\ps@draft{\let\@mkboth\@gobbletwo
\def\@oddhead{}
\def\@oddfoot
{\hbox to 7 cm{$\scriptstyle Draft\ version:\ \draftdate$ \hfil}
\hskip -7cm\hfil\rm\thepage \hfil}
\def\@evenhead{}\let\@evenfoot\@oddfoot} \catcode`\@=12
\documentstyle[12pt]{article}
\newcommand{\BE}{\begin{eqnarray}}
\newcommand{\EN}{\end{eqnarray}}
\newcommand{\be}{\begin{equation}}
\newcommand{\en}{\end{equation}}
\newcommand{\non}{\nonumber}
\newcommand{\no}{\noindent}
\newcommand{\vs}{\vspace}

\newcommand{\p}{\partial}
\newcommand{\Bbb}{\bf}

\begin{document}

\title{Universality in Blow-Up for Nonlinear Heat
Equations}

\author{J.Bricmont\thanks{Supported by EC grant SC1-CT91-0695}\\UCL,
Physique Th\'eorique,
Louvain-la-Neuve, Belgium\and
A.Kupiainen\thanks{Supported by NSF grant DMS-8903041} \\Helsinki
University, Mathematics Department,\\ Helsinki, Finland}

\date{}

\maketitle
\begin{abstract}
We consider the classical problem of the blowing-up of
solutions of the nonlinear heat equation. We show that there exist
infinitely many profiles
around the blow-up point, and for each
integer $k$, we construct a set of codimension
$2k$ in the space of initial data
giving rise to solutions that blow-up according to the given profile.

\end{abstract}
\section{Introduction}
\setcounter{equation}{0}

We consider the problem of the blow-up of solutions of the initial value
problem
\be
u_t = u_{xx} + u^p
\en
where $p > 1, u=u(x,t),x \in \Bbb R$, and $u(\cdot,0)=u_0\in
C^0({\bf R})$. It is well-known
that, for a large
class of initial data $u_0$, the solution will diverge
in a finite time at
a single point (for
reviews on this problem, see \cite{HV,L}).

We are interested in the profile
of the solution at
the time of blow-up. To explain what this means,
let us fix the blow-up
point to be 0 and the
blow-up time to be $T$. Then, we ask whether it
is possible to find a
function $f^*(x)$ and a
rescaling $g(t,T)$ so that
\be
\lim_{t \uparrow T} (T-t)^{\frac{1}{p-1}} \;
u(g(t,T)x,t)=f^*(x)
\en
Moreover, we want to see how $g$ or $f^*$
depend on the initial data.

The
prefactor
$(T-t)^{\frac{1}{p-1}}$ in (2) can be understood easily:
for initial data $u_0(x)$ constant in $x$, $u(t)$ solves
the ODE $\dot u = u^p$, i.e. $u(t) =
((p-1)(T-t))^{\frac{1}{1-p}}$ for $T=(p-1)^{-1}u_0^{1-p}$.
We therefore expect that
$f^*(0)=(p-1)^{\frac{1}{1-p}}$. However, we want to obtain $f^*$ for $x
\neq 0$. In \cite{HV1,HV2,HV3,V} (see also \cite{F1,F2})
several possible $f^*$'s are discussed, and the set of initial data that
will lead to a given $f^*$ is partially characterised.
In the present paper, we shall show
that there exists, in the space of initial
data $C^0({\bf R})$,  sets ${\cal M}_k$ of codimension $2k$,
such that, for $u_0 \in {\cal M}_k$,
the limiting
behaviour (2) is obtained, in the case $k=1$, for
\BE
g(t,T)&=&((T-t) | \log (T-t)|)^{\frac{1}{2}}\\
f^*(x)&=&\left(p-1+
b^*x^2\right)^{\frac{1}{1-p}}
\EN
where $b^*= \frac{(p-1)^2}{4p}$, and in the case $k>1$ for
\BE
g(t,T)&=&(T-t)^{\frac{1}{2k}}\\
f_b^*(x)&=&\left(p-1+
bx^{2k} \right)^{\frac{1}{1-p}}.
\EN
where now $b$ is an arbitary positive number.

The lowest codimension 2 corresponds to fixing two parameters
in the data that specify the blow-up point and time. To reach
the other "strata", $2k-2$ additional parameters describing the data
need to be fixed.

Now we want to relate this problem to the
 renormalization group approach to
the study of the
asymptotic behaviour of solutions of
nonlinear partial differential
equations; this approach was
initiated and developed in \cite{Ba,Go1,Go2}
and, from a  mathematical
point of view, in
\cite{BKL}. Although our actual proofs do not rely very much on this
approach, the ideas used
here are close to the ones of the
renormalization group. In this approach,
the long-time
behaviour of the solution is related to
the existence of fixed points of
the renormalization group
transformation (which basically amounts to
solving the PDE over a finite
time interval, and
combining this with some scaling transformations). A given
asymptotic behaviour is obtained,
provided the initial data lie in the basin of
attraction of a given fixed
point. This basin of
attraction is called a ``universality class".
In the simplest cases, the
fixed points are stable
but, in general, there can be one or more
unstable or neutral directions
for the renormalization group flow around the
fixed point. This is exactly what happens
here: $f^* $
and $f^*_b$ can be viewed as
a fixed points of a renormalization group
transformation having $2k$ unstable (``relevant", in renormalization
group terminology)
directions. Thus, to converge towards the
fixed point, one has to fine-tune
$2k$ parameters (one
for each unstable direction) and this explains
why ${\cal M}_k$ is of
codimension $2k$, and in what
sense $f^*$ is ``universal". In addition,
we encounter also one neutral
(``marginal") mode, which, for $k=1$,
turns out to be stable when nonlinear effects are taken into account
and for $k>1$ parametrizes a curve of fixed points. Thus,
our result is also
connected to the center manifold theory.

Our results are perturbative, i.e. the sets ${\cal M}_k$ consist of initial
data that are close to the corresponding fixed point. Therefore, our
results are similar to those of Bressan \cite{Br} who considers a
nonlinearity $e^{u}$
instead of $u^p$ and obtains the universal profile analogous
to our $k=1$ case. However, his method is
different from ours and we
obtain a control over the limit (2) which is uniform in $x$.

\vs{3mm}

To describe our main results, let us fix a positive number $T$, and
introduce
\be
 f(\xi)=\left\{ \begin{array}{ll} T^{1\over
p-1}u_0(T^{1\over 2k}\xi)& k>1\\
T^{1\over
p-1}u_0((|\log T|T)^{1\over 2}\xi)& k=1
\end{array}\right.
\en
Let first $k=1$. We write the initial data as
\be
f(\xi) = f^*(\xi)(1 + {d_0+d_1\xi\over
p-1 + b^*\xi^2}) +
g(\xi)
\en
where, given $g\in C^0({\bf R})$, $d_0$ and $d_1$ are the two
parameters to be fixed. We have

\vs{3mm}

\no {\bf Theorem 1.}
{\it There exists a $T_0 > 0$ such that, for each $0 < T
< T_0$
and $g \in C^0({\Bbb R})$ with $\Vert g\Vert_{\infty} < (\log\ T_0)^{-2}$
one
can
find $d_0$ and $d_1$, such that the equation} (1) {\it with
the initial data} (7),(8) {\it has a unique classical solution
$u(x,t)$ on ${\Bbb R} \times
[0,T)$ and
\be \lim\limits_{t\uparrow T}\ (T-t)^{1\over p-1}
u(((T - t) | \log\ (T - t)|)^{1\over 2}\xi,t) = f^*(\xi)
\en
uniformly in $\xi$ on} ${\Bbb R}$.

\vs{3mm}

\no{\bf Remark 1.} We get a much more detailed information
on $u(x,t)$, see the Proposition in Sect.3.

\vs{3mm}

\no {\bf Remark 2.} To leading order, $d_0(p-1)^{-{p\over{p-1}}}
 = -\frac{a}{\log T}$, with
$a=2b^*
(p-1)^{\frac{2p-1}{1-p}}$, and is independent of $g$ (see Lemma 2 in
Sect.3
 below). This means that $d_0$ (and $d_1$) are nonzero, even if we have
$g(\xi)=0$
in (8).

\vs{3mm}

\no{\bf Remark 3.}
The proof can be extended to more general
nonlinearities, than (1): we actually give below the proof for
the equation
\be
u_t = u_{xx} + u^p + F(u)
\en
and we will assume that $F: {\Bbb R} \to {\Bbb R}$ is Lipschitz and
satisfies
\be
|F(u)| \leq C(1 + |u|^q)
\en
with $0 \leq q < p$,
and
$$
|F(\lambda u_1) -F(\lambda u_2)| \leq C \lambda^q |u_1 -u_2|
$$
for $|u_1|, |u_2| \leq 1$ and $\lambda \geq 1$.
With little extra effort the proof extends to
nonlinearities
$F(u,u_x)$ in (1.10) that depend on $u_x$. We need then
\be
|F(u,u_x)| \leq C(1 + |u|^q + |u_x|^r)
\en
with $q$ as in (11),
\be
r < {2p\over p + 1},
\en
and the corresponding Lipschitz bound.
\vs{3mm}

\no{\bf Remark 4.} The proof also extends to
 $x \in {\Bbb R}^n,\ n > 1; d_1\xi$ in
(8) becomes $\vec d_1 \cdot \vec
\xi$ where $\vec d_1$ is
an $n$-component vector and $\xi^2$ becomes $\| \vec
\xi \|^2$. Thus, we need to
fix $n+1$ coefficients. For simplicity, we shall keep $n = 1$.

\vs{3mm}

For $k>1$, we take the data of the form
\be
f(\xi) = f_b^*(\xi)(1 + \sum\limits_{i=0}^{2k-1} d_i\xi^i
(p-1 + b\xi^{2k})^{-1}) +
g(\xi)
\en
where $d_i$ are now the parameters to be fixed, once a $g\in
C^0({\bf R})$ is given. We have then the

\vs{3mm}

\no{\bf Theorem 2.} {\it
There exist  $T_0 > 0$  and  $\varepsilon > 0$ such that\ for
$0<T< T_0$  and  $g$  in  $C^0({\Bbb R})$  with
$
\Vert g\Vert_{\infty} < \varepsilon$
there are constants  $d_i \in {\Bbb R}$  such that\ the equation}
(1){\it with the
initial data} (14) {\it has a unique classical solution
$u(x,t)$ on ${\Bbb R} \times
[0,T)$ and}
\be \lim\limits_{t\uparrow T}\ (T-t)^{1\over p-1}
u((T - t) ^{1\over 2k}\xi,t) = f_{b^*}^*(\xi)
\en
{\it uniformly in $\xi$ on ${\Bbb R}$,
for some  $b^* > 0$, where  $b^* \to b$  as  $\varepsilon \to 0$  and}
$T_0 \to 0$.

\vs{3mm}

\no {\bf Remark 1}.
Thus we have, for  $k > 1$, a line of fixed points  $f_b^*$
and given initial data in the codimension  $2k$  set in $C^0({\bf R})$,
the  $u(x,t)$  arrives to this line as  $t \to T$.
The only effect of the  $g$  in the data (14) is to ''renormalize'' the
$b$  occurring in the data.  Compare with the  $k = 1$  case where there
was a unique fixed point  $f^*$.

\vs{3mm}

\no {\bf Remark 2}. Note that our assumptions, in both Theorems,
allow initial data that are
not everywhere positive.

\vs{3mm}

\no {\bf Remark 3}.
Again, more general equations can be treated, but we leave that
formulation for the reader.

\section{ Dynamical systems formulation}

\setcounter{equation}{0}

In this section we describe a change of variables that transforms
the problem (1.10) into a problem of long time asymptotics. We also
explain the main ideas of our proof.

We write (1.10) in the ``blow--up--variables'':
given a $u: {\Bbb R} \times [0,T) \to {\Bbb R}$,
define
$\varphi : {\Bbb R} \times [-\log\ T,\infty) \to
{\Bbb R}$ by
\be
u(x,t) = (T - t)^{-{1\over
p-1}}\varphi ({x\over (T - t)^{1/2k}}, -\log\ (T - t)).
\en
Then $u$ is a classical solution of (1.10)
if and only if $\varphi (\xi,\tau)$ is a classical solution of
\BE
\dot {\varphi}& =& L_\tau^{-2}
\varphi'' - {1\over 2k}\xi \varphi' - {1\over p -
1}\varphi +
\varphi^p + F_{\tau}(\varphi)\\ \varphi (\xi,\tau_0)& =& T^{1\over
p-1}u_0(T^{1\over 2k}\xi)
\EN
where $\tau_0=-\log T$,
\be
L_{\tau} = e^{{1\over 2}\tau (1-1/k)}.
\en
and
\be
F_{\tau}(\varphi) = e^{-{p\over p-1}\tau}F(e^{\tau\over p-1}\varphi). \en
We will construct global solutions of (2), with
suitable initial data thereby establishing blow--up for (1.10).
Note that, for $k=1$, the scaling in (1) differs from the one used in (1.9)
by a
factor $\tau^{1/2}$.

\vs{3mm}

Consider first the $k=1$ case with $F=0$.
To understand the dynamics of (2), let us start by considering its
linearization around the constant solution $\varphi=
(p-1)^{1\over 1-p}$. The linear problem is $\dot {\phi}=
{\cal L}\phi$, where
\be
{\cal L} = \partial^2 - {1\over 2}\xi \partial + 1 \; .
\en
with $\partial=\partial_{\xi}$.
 Hence, the first thing we have to do to understand the
stability of the constant solution is to study the
spectrum of the linear operator ${\cal L}$.

${\cal L}$ is self--adjoint on ${\cal D}({\cal L})
\subset L^2({\Bbb R},d\mu)$ with
\be
d\mu (\xi) = \frac{e^{-\xi^2/4}d\xi}{\sqrt{4 \pi}}
\en
The spectrum of ${\cal L}$  is
\be
spec({\cal L}) = \{1 - {n\over 2}\mid n \in {\Bbb N}\}
\en
and we take as eigenfunctions multiples of Hermite polynomials \be
h_n(\xi) = \sum\limits_{m=0}^{[{n\over 2}]}{n!\over m!(n-2m)!}(-1)^m
{\xi}^{n-2m} \en
that satisfy
\be
\int h_nh_md\mu = 2^n n!\delta_{nm}
\en
and
\be
{\cal L}h_n = (1 - {n\over 2})h_n.
\en
Thus the derivative of the RHS of (2) at the constant solution
has $2$
expanding
(``relevant'') directions and one neutral (``marginal'') one,
$h_{2}=\xi^2-2$.

How do we understand now the emergence of the fixed point
$f^*$?
We get a clue on what should happen
by considering the following scaling: let
\be
\varphi_L(\xi,\tau) = \varphi (L\xi,L^{2}\tau)
\en
$\varphi_L$ satisfies the equation
\be
H_1(\varphi_L)
  = L^{-2}(-\dot {\varphi}_L + \varphi''_L) +
F_{L^{2}\tau}(\varphi_L).
\en
where we defined
\be
H_k(\varphi)= {1\over 2k}\xi \p_\xi \varphi
+ {1\over p-1}\varphi -
\varphi^p\; .
\en
Hence, as $L \to \infty$,
we expect the solutions of
$H_k(\varphi) = 0$ (for $k=1$)
to be relevant. These are, for any $k$,
given by the one-parameter family
$f_b^*$ (see (1.6)). Before we explain why only one $b^*$ is
selected, we will compare the above with the $k>1$ case.

\vs{3mm}

For $k>1$,
as $\tau\to\infty$, we expect
the solution of
\be
\dot {\varphi} = -{1\over 2k}\xi \varphi' - {1\over p-1}
\varphi + \varphi^p\en
to be relevant (see (2,4)).  This can of course be integrated in closed
form, but before doing that, let us first look at
its linearization around the constant solution $\varphi=
(p-1)^{1\over 1-p}$. The linear problem is $\dot {\phi}=
{\cal L}_\infty\phi$, where
\be
{\cal L}_\infty = - {1\over 2k}\xi \partial + 1 \; .
\en
and so, e.g. in the space of polynomials, we have now
$2k$ expanding directions corresponding to
$\xi^n$, for $n<2k$.

Equation (15) is solved by putting
$\varphi (\xi,\tau) = e^{-{\tau\over p-1}}h(e^{-\tau/2k}\xi,\tau)$
whereby  $\partial_{\tau}h(y,\tau) = e^{-\tau}h(y,\tau)^p$  and so, for
$\rho = \tau - \tau_0$
\be
\varphi (\xi,\tau) = {e^{-{\rho\over p-1}}f(e^{-\rho/2k}\xi)\over
[1 - (p-1)f(e^{-\rho/2k}\xi)^{p-1}(1 - e^{-\rho})]^{1/p-1}}
\en
where  $\varphi (\xi,\tau_0) = f(\xi)$.  Depending on  $f$, (17) has
several possible asymptotics as  $\rho \to \infty$.  In the space of
constant $f$'s we have the stable  $f = 0$  and unstable  $f =
(p - 1)^{-{1\over p-1}}$  fixed points.  The latter is stable in a
suitable codimension  $2k$  space: let us consider say  $f$  smooth,
\be
f(0) = (p - 1)^{-{1\over p-1}},\ f^{(\ell)}(0) = 0\ \ \ell < 2k,\
f^{(2k)}(0) = \beta < 0\en
and
\be
0\leq f(\xi) < (p - 1)^{-{1\over p-1}}\ \;\; \xi \neq 0.\en
Then, for all $\xi\in{\bf R}$
\be
|\varphi (\xi,\tau) - f_b(\xi)|
\mathrel{\mathop{\longrightarrow}\limits_{\tau \to \infty}} 0\en
where
\be
f_b(\xi) = (p - 1 + b\xi^{2k})^{-{1\over p-1}}\en
for  some $b$ depending on $\beta$,$ k$, $p$.

These considerations thus lead us to expect (2) to
have global solutions with initial data in a suitable codimension  $2k$
set in a ball around  (21)  in a suitable Banach space.
Note however, that the perturbation  $L^{-2}_{\tau}\varphi_{\xi\xi}$  in
(2) is a very singular one: we certainly need to keep track of its
smoothing effects.  On the other hand, we want to retain as much
as possible of the nice picture obtained above in the $\tau\to\infty$
limit. We explain now how this is done for (2) linearized
around the  constant solution, leaving the nonlinear analysis to
the actual proof in Section 4.

The linearization of (2) around the constant solution is
$
\dot {\phi}=
{\cal L}_{\tau}\phi
$
where
\be
{\cal L}_{\tau} = L_{\tau}^{-2}\partial_{\xi}^2 - {1\over 2k}
\xi \partial_{\xi} + 1,
\en
In order to study linear stability, we thus
need some properties of the fundamental solution
$K_{\tau \sigma}$  of (22), i.e.
\be
\partial_{\tau}K_{\tau \sigma} = {\cal L}_{\tau}K_{\tau \sigma},\quad
K_{\sigma \sigma} = {\rm id}.
\en
$K_{\tau \sigma}$ is conveniently found, by conjugating
the problem (23) to a time independent one:
\be
K_{\tau \sigma} = S_{\tau}e^{(\tau - \sigma){\cal L}}S_{\sigma}^{-1}
\en
where
\be
(S_{\tau}\theta)(\xi) = \theta (L_{\tau}\xi)\en
and ${\cal L}$  is given in (6).  Thus, in terms of kernels
\be
K_{\tau \sigma}(\xi,\xi') = L_{\sigma}e^{(\tau - \sigma){\cal L}}
(L_{\tau}\xi ,L_{\sigma}\xi')
\en
and, since the  kernel of $e^{\rho{\cal L}}$ is given
explicitely by Mehler's formula \cite{Si}:
$$
e^{\rho {\cal L}}(\xi,{\xi'}) = [4\pi (1 - e^{-\rho})]^{-1/2}e^{\rho} \exp\
[-{(\xi e^{-\rho/2} - {\xi'})^2\over 4(1 - e^{-\rho})}],
$$
(26) can be written in the form
\be
K_{\tau \sigma}(\xi,\xi')
= e^{\rho}\delta_{L^2}(e^{-\rho /2k}\xi - \xi')
\en
where  $\rho = \tau - \sigma$,
$L^2 = L^2_{\sigma}(1 - e^{-\rho})^{-1}$,
and
\be
\delta_{L^2}(\xi) = {L\over \sqrt {4\pi}}e^{-L^2\xi^2/4}
\en
(27) and (28) show clearly that the effect of the
$L_{\tau}^{-2}\partial_{\xi}^2$  is to smoothen the kernel of the linear
problem (see (16)),  $\dot {\varphi} = (1 - {1\over 2k}\xi
\partial_{\xi})\varphi$,  which is just
\be
K_{\rho}^{\infty}(\xi,\xi') \equiv e^{\rho}\delta(e^{-\rho /2k}\xi -
\xi')
\en
i.e.\ the distributional limit of (27) as  $L \to \infty$  i.e.\ as
$\sigma \to \infty$.

As in the $k=1$ case, we may now study the stability of the
linearization
in a Hilbert space. The ''eigenfunctions'' of  $K_{\tau \sigma}$
are readily obtained
 From (9) by the conjugation (24).  We put
\be
h_n(\xi,\tau) = L_{\tau}^{-n}h_n(L_{\tau}\xi) =
\sum\limits_{m=0}^{[{n\over 2}]}{n!\over m!(n - 2m)!}(-L^{-2}_{\tau})^m
\xi^{n-2m}\en
(note that  $h_n \to \xi^n$  as  $\tau \to \infty)$  whereby
$h_n(\cdot,\tau)$  form a basis of  $L^2({\Bbb R},d\mu_{\tau})$
where
\be
d\mu_{\tau}(\xi) = {L_{\tau}\over \sqrt {4\pi}}e^{-L^2_{\tau}\xi^2/4}
d\xi\en
and
\be
(h_n(\cdot,\tau),h_m(\cdot,\tau))_\tau =
\int h_n(\xi,\tau)h_m(\xi,\tau)d\mu_{\tau}(\xi) = L_{\tau}^{-2n}2^nn!
\delta_{nm}.\en
We then have
\be
K_{\tau \sigma}h_n(\cdot,\sigma) = e^{(\tau - \sigma)(1-n/2k)}
h_n(\cdot,\tau)\en
which should be compared with  $K_{\rho}^{\infty}p_n =
e^{\rho(1-{n\over 2k})}p_n$  for  $p_n(\xi) = \xi^n$.
The $h_n$ with $n<2k$ form thus a convenient basis for
the expanding modes.

\vs{3mm}

Finally we want to comment on the effect of the nonlinear terms.
The linear analysis presented above deals with deviations from the
constant solution and turns out to describe the solution well for
$|\xi|$ not too large. We thus need to understand why the fixed points
$f^*_b$ are selected, and, for $k=1$, why only one $b^*$ occurs. Finally
we need to understand the stability problem for $|\xi|$ large. We shall
only
discuss here the $k=1$ case, since $k>1$ is actually easier (see Section
4).

Consider $k=1$. We introduce
\be
\varphi_b(\xi,\tau) = (p - 1
+ b\xi^{2}/\tau)^{1\over 1-p}
\en
where the factor $\tau$ can be understood by comparing the scaling in (1)
and
in (1.9), and we
study the flow near $\varphi_b$. Let
us
rewrite (2) in terms of $\eta$, where
\be \varphi (\xi,\tau) =
\varphi_b(\xi,\tau) + \eta (\xi,\tau).
\en
We get, using $H_1(\varphi_b) =
0$ (see (14)),
\BE
\dot {\eta}& =& \eta'' - H_1(\varphi_b + \eta) + H_1(\varphi_b)
 + \varphi''_b - \dot {\varphi}_b + F_{\tau}(\varphi_b + \eta) \non\\
&=& ({\cal L} + W)\eta + M(\eta) + \varphi''_b - \dot
{\varphi}_b + F_{\tau}(\varphi_b + \eta)
\EN
where we write $\frac{-1}{p-1}=1-\frac{p}{p-1}$, and introduce
\BE
&&W = p(\varphi_b^{p-1} - {1\over p - 1})\\ &&M(\eta) =
(\varphi_b + \eta)^p - \varphi_b^p - p\varphi_b^{p-1}\eta.
\EN
 $\cal L$, given by (6), has two unstable modes.
Note that, formally, (i.e., for $\xi$ of order one) $W$ is ${\cal
O}(\tau^{-1}),\ M$ is
nonlinear  in $\eta$ and $\varphi''_b - \varphi'_b$ is ${\cal
O}(\tau^{-1})$.
Our goal will be
to construct a center manifold for (36), i.e.\ to find the parameters
$d_0,d_1$ in (1.8),
such that the flow of (36) stays bounded.

To explain the idea of the proof we first consider the special case $p = 2$
and $\eta$ even in $\xi$, which will imply $d_1=0$ in (1.8). This example
contains all the
relevant features of the general case. Now,
\be
\varphi_b(\xi,\tau) = (1 + b\xi^2/\tau)^{-1} .
\en

It is convenient to first find
$d_0$
approximately, exact to order $1/\tau$. Let
\be
\eta_0 (\tau) = \frac{a}{\tau}
\en
and define $\psi$ by
$$
\eta = \eta_0 + \psi.
$$
Then $\psi$ satisfies the equation
\be
\dot {\psi} = ({\cal L} + V)\psi + N(\psi) + \alpha \en
with (for later purpose we write this for general $p$)
\BE
V& =& p((\varphi_b + \eta_0)^{p-1} - {1\over p - 1})\\
N(\psi)& =& (\varphi_b + \eta_0 + \psi)^p - (\varphi_b + \eta_0)^p -
p(\varphi_b +
\eta_0)^{p-1}\psi + F_{\tau}(\varphi_b + \eta_0 + \psi)\\
\alpha& = &\varphi''_b - \dot {\varphi}_b +
({\cal L} + W)\eta_0 - \dot {\eta}_0 + M(\eta_0) \nonumber \\
& =& \varphi''_b-\dot \varphi_b + \eta_0+ W \eta_0 - \dot \eta_0 + M
(\eta_0).
\EN

We shall see how to choose $a$ and $b$ so that the flow of $\psi$ in (41)
can remain bounded.
Let us  decompose $\psi$ as \be \psi
 = \psi_0(\tau) + \psi_2(\tau)h_2 +
\psi^{\perp} \en where $\psi^{\perp}$ is orthogonal
 to $h_n,\ n \leq 2$
(we will later in the actual proof refine (45)).
Next we expand $V$ and $\alpha$ (for $\xi={\cal O}(1))$:
\BE
&&V = -{2b\xi^2\over \tau} + {2a\over \tau} + {\cal O} ({1\over \tau^2})
\\
&&\alpha =(a-2b)\tau^{-1} +(a+a^2+ (12b^2 -b - 2ab)\xi^2))\tau^{-2}
+ {\cal O}(\tau^{-3}).
\EN
Inserting (40), (45) in (41) and
retaining only the leading terms in
$1/\tau$ and $\psi_i,\ i = 0,2$, we get from $\dot
{\psi}_i = (2^i i!)^{-1}(h_i,\dot{\psi})$
($(\cdot,\cdot)$ is the scalar product of $L^2({\Bbb R},
d\mu)$):
\BE
\dot {\psi}_0 &=& \psi_0 +
(a - 2b)\tau^{-1} + R_0\\ \dot {\psi}_2 &=& \beta
\tau^{-1}\psi_2 + (12b^2 -b - 2ab)\tau^{-2} +
R_2
\EN
where $R_0 = {\cal O}(\tau^{-2} + \tau^{-1}|\psi| + |\psi|^2),
 R_2 = {\cal O}(\tau^{-3} + \tau^{-1}|\psi_0| + \tau^{-2}|\psi_2| +
|\psi|^2)$,and
$\beta = 2a - {1\over 4}b(\xi^2h_2,h_2) = 2a - 20b$ (coming from the $V
\psi$ term in (41)).
We choose now $a$ so that the ${\cal O}(\tau^{-1})$ term in
$\dot {\psi}_0$ vanishes i.e.
\be
a = 2b
\en
and $b$ such that  the ${\cal O}(\tau^{-2})$ term in $\dot {\psi}_2$ is
zero:
\be
b = b^*=1/8.
\en
Note that this choice correspond to $b=b^*$ in (1.4) for $p=2$ and $a$ as
in Remark 1.2.
Then $\beta=-2$ and our equations read
\be
\dot {\psi}_0 = \psi_0 + R_0,\ \ \dot {\psi}_2 = -{2\over \tau}\psi_2 +
R_2.
\en
Now,
\be
\psi_0 = {\cal O}(\tau^{-2}),\ \ \psi_2 = {\cal O}
((\log\ \tau)\tau^{-2}) \en
would  be consistent solutions. Of course,
we need to show that the expanding variable $\psi_0$
will satisfy (53) by a suitable choice of
$\psi_0(\tau_0)$, i.e. of the parameter $d_0$ in Theorem 1. This is rather
easy to do, using the fact
that $\psi_0$ is expanding; in the general case (with $d_1 \neq 0$), we
shall use a topological
argument.

If we were to expand $\psi^{\perp}$ in (45) as
\be
\psi^{\perp} = \sum\limits_{i=2}^{\infty} \psi_{2i}h_{2i} \en
we would then formally get
\be
\dot {\psi}_{2i} = -(i - 1)\psi_{2i} + {\cal O}(1/\tau^{1+i}) +
N(\psi)_{2i}
\en
(in $\alpha$, we have an extra factor of $\tau^{-1}$ coming from the
derivatives or from $\eta_0$)
and the formal solution would be
\be
\psi_{2i}(\tau) = {\cal O}(\tau^{-1-i})
\en
so that $\psi_{2i}(\tau)h_{2i}(\xi\tau^{1/2}) \to 0$
as $\tau \to \infty$, for all $i$ (to prove (1.9), we need to scale $\xi$
here by $\tau^{1/2}$, see (1)). However, (54) will not be a
good representation for
large
$\xi$ and we need to proceed differently.

We decompose $\psi$ to a part localized on an interval around
the origin and to a part describing the large $|\xi|$ behaviour.
For this let $\chi\in C_0^{\infty}({\Bbb R})$ be non-negative,
$\chi = 1$ on $[-1,1]$ and $\chi = 0$ on $[-2,2]^c$. Let $K > 0$, and put
\be
\chi(\xi,\tau) = \chi ((K\tau^{1/2})^{-1}\xi).
\en
$K$ will be taken suitably large, see below. Let now
\be
\psi = \psi\chi + \psi(1 - \chi) \equiv
\psi_s + \psi_l
\en
The ``small $\xi$ part'',
$\psi_s$, will be decomposed as above:
\be
\psi_s(\xi,\tau) = \psi_0(\tau) + \psi_2(\tau)h_2(\xi) +
\psi^{\perp}(\xi,\tau)
\en
and we shall prove that
\BE
|\psi_0(\tau)|& \leq& C\tau^{-2}
\\
|\psi_2(\tau)|& \leq& \tau^{-2+\delta}\\
|\psi^{\perp}(\xi,\tau)| &\leq& C(1 + |\xi|^3)\tau^{-2}\\
\Vert \psi_l (\cdot,\tau)\Vert_{\infty} &\leq& C\tau^{-1/2}.
\EN
for any $\delta>0$. This bound on $\psi_2$ is a convenient upper bound on
the $(\log \tau)
\tau^{-2}$ behaviour which is expected on the basis of (52,53). Note that
$\psi_0, \psi_2$ are functions only of $\tau$, while $\psi^{\perp}, \psi_l$
depend on $\tau$ and $\xi$.

The detailed $\psi_l$ bound will be explained in the
proof below, but here we want to comment only on the
decay in $\xi$ that we expect. The reason
that (54) is not a good expansion is that the eigenfunctions
of ${\cal L}$, i.e.\ $h_i$, grow at
infinity; the more they are contracted by $e^{\tau{\cal L}}$, the more they
grow. This would
make the nonlinear term in (41) impossible to control. However, for $|\xi|
> K\tau^{1/2}$, the
$V$ in (42) (see (39)) is not any more small; actually ${\cal L} + V$
behaves like ${\cal L} -
{ p\over p-1}$ in that region and this operator has purely negative
spectrum. This is why an
$L^{\infty}$--bound such as (63) will hold.

The proof of the general $p$ case is very similar. We have now $2$
expanding
modes (if $\eta$ is not even), and the number $b$ is again determined from
a condition
that the neutral mode contracts like $\tau^{-2}$ (with possibly logarithmic
corrections).

\section{ The proof, $k=1$}

\setcounter{equation}{0}

Theorem 1 reads, in terms of $\varphi$, as

\vs{3mm}

\no {\bf Theorem 3}. {\it There exists a $T_0 > 0$
such that, for each $0 < T
< T_0$
and $g \in C^0({\Bbb R})$ with $\Vert g\Vert_{\infty}
< (\log\ T_0)^{-2}$ one can
find $d_0$ and $d_1$, such that the equation} (2.2) {\it with
the initial data} (2.3, 1.7, 1.8) {\it has a unique classical solution
$\varphi (\xi,\tau)$ satisfying}
$$
\lim\limits_{\tau \to \infty}\
\Vert \varphi (\cdot\tau^{1/2},\tau) - f^*(\cdot)\Vert_{\infty} = 0
$$

\vs{3mm}

We consider the equation for $\psi$ given by (2.41)-(2.44). The initial
data is given by
(see (2.3),(1.7),(1.8),(2.35),(2.40))
\be
\psi(\xi,\tau_0)=
\varphi_{b^*}(\xi,\tau_0)( {d_0+d_1\xi\tau_0^{-1/2} \over
p-1 + b^*\xi^2/\tau_0})-a\tau_0^{-1} +
g(\xi\tau_0^{-1/2})
\en
Next we state the properties of $\psi$ that we want to establish. We
write $\psi$ as in (2.58),
\be
\psi=\psi_s+\psi_l
\en
with this time
\be
\psi_s(\xi,\tau) = \sum\limits_{m=0}^{2} \psi_m(\tau)h_m(\xi) +
\psi^{\perp}(\xi,\tau). \en We will prove the

\vs{3mm}

\no {\bf Proposition.} {\it With the assumptions of the
Theorem, for any $\delta > 0$, there exist a $\tau_0$ and constants
$d_0, d_1$, such that $ \psi$, }given by (2),(3) {\it will satisfy
\be
|\psi_m(\tau)| \leq \left\{ \begin{array}{ll} A\tau^{-2}& m = 0,1\\
\tau^{-2+\delta}& m=2
\end{array}\right.
\en
\be
|\psi^{\perp}(\xi,\tau)| \leq A(1 + |\xi|^{3})\tau^{-2} \en
and
\be
\Vert \psi_l\Vert_{\infty} \leq A_l \tau^{-{1\over 2}}. \en
for some constants $A$, $A_l$, uniformly on $[\tau_0,\infty)$.}

\vs{2mm}

\no{\bf Remark.} Theorem 3 follows
immediately from the Proposition, which implies that
$\Vert \psi(\cdot \tau^{1/2},
\tau)\Vert_{\infty}\leq{\cal O}(\tau^{-1/2 })$ (by (2.57), $|\xi|$
in $\psi_s$
is bounded by $2K\tau^{1/2 }$). The $\delta$ in (4)
may be made arbitary small by increasing
$\tau_0$ (i.e. decreasing the data in $\psi$ or, equivalently, taking $T$
small and $u_0$
large in (2.3)). It will be convenient in the proof to distinguish between
$A$ and $A_l$.

\vs{2mm}

\no{\bf Proof.} Let us assume that (4)-(6) hold for some
$\sigma \geq\tau_0$
and study the existence and properties of the solution
for subsequent times on an interval $ [\sigma,\sigma+\rho]$ .
 We shall choose below a sufficiently large constant
$\rho$, and prove
iteratively our bounds on intervals of the form $[ {\tau}_{n},
{\tau}_{n+1}]$ with ${\tau}_{n}= {\tau}_{0}+ n \rho$.

To prove existence and uniqueness, write (2.41) as an integral equation \be
\psi (\tau) = K(\tau,\sigma)\psi(\sigma) +
 \int\limits_{\sigma}^{\tau} ds K(\tau,s)[N(\psi (s)) + \alpha (\cdot,s)]
\en
for $\psi (\tau) \equiv \psi (\cdot,\tau)$. $K$
is the fundamental solution of the linear equation $\dot{K}=({\cal L}+V)K$.
We study the three terms in (7) separately.

We expand the linear term in $\psi$ as in (2) and (3): \be
K(\tau,\sigma)\psi(\sigma) = \sum\limits_{m=0}^{2} \theta_mh_m +
\theta^{\perp}
+ \theta_l.
\en
with
\be
\theta_l(\xi,\tau)=(1-\chi(\xi,\tau))(K(\tau,\sigma)\psi(\sigma))(\xi). \en
Lemma 1 collects the bounds for the $\theta$'s:

\vs{3mm}

\no {\bf Lemma 1}. {\it For any $\rho>0$, there exists a $\tau_0$ such
that, if $\psi(\sigma)$
satisfies} (4)-(6) {\it for $\sigma \geq \tau_0$, then, for $\tau \leq
\sigma+\rho$,}
\BE
|\theta_m(\tau)-e^{(1-{m\over 2})(\tau-\sigma)} \psi_m(\sigma)|&\leq&
(\tau-\sigma)C\tau^{-3+\delta} \;\;\;m=0,1\\
|\theta_{2}(\tau)-({\sigma\over\tau})^2\psi_{2}(\sigma)|&\leq &
(\tau-\sigma)CA\tau^{-3}\\
|\theta^\perp(\xi,\tau)|&\leq& C(e^{-{1\over
2}(\tau-\sigma)} A+ e^{-(\tau-\sigma)^2}A_l)(1+|\xi|^{3}) \tau^{-2}\\
\|\theta_l(\cdot,\tau)\|_\infty &\leq& C(A_l e^{-{(\tau-\sigma)\over p}}+A
e^{(\tau-\sigma)})
\tau^{-{1\over 2}}
\EN

\vs{3mm}

Here and below we use $C$ or $c$ to denote a generic constant, which may
vary from place to
place. $C$ may depend on $K$ in (2.57), but not on $A$, $A_l$ or
anything else (unless explicitely stated
otherwise), and, since we
shall consider $K$ as fixed, but sufficiently large, these constants are
fixed
also.

For the
$\alpha$-term, we need to specify $\eta_0$, i.e.  the number $a$ in (2.40),
as well as
$b^*$, so that the contribution of $\alpha$  to $\psi$ is (almost) of
the same order of magnitude as the bounds (10)-(13). We have

\vs{3mm}

\no{\bf Lemma 2}. {\it Let }
\be
a=2b^* (p-1)^{\frac{2p-1}{1-p}} = \frac{(p-1)^{\frac{1}{1-p}}}{2p}
\en
{\it
i.e. $b^*= \frac{(p-1)^2}{4p}$ and set $b=b^*$ in (2.44).
Define
$$
{\cal A}(\xi, \tau, \sigma)=
\int\limits_{\sigma}^{\tau} ds K(\tau,s) \alpha (\cdot,s)
$$
Then ${\cal A}(\xi, \tau, \sigma)$ has  an expansion as in} (8):
\be
 {\cal A}= \sum\limits_{m=0}^{2} {\cal A}_m h_m + {\cal A}^{\perp} + {\cal
A}_l
\en
{\it with}
\BE
|{\cal A}_m(\tau, \sigma)|
&\leq& (\tau-\sigma) Ce^{(\tau-\sigma)}\tau^{-2}\;\;\;m=0,1\\
|{\cal A}_{2}(\tau, \sigma)|&\leq& (\tau-\sigma)C\tau^{-3}\\
|{\cal A}^\perp(\xi,\tau, \sigma)|&\leq& (\tau-\sigma)C(1+|\xi|^{3})
\tau^{-2}\\
\|{\cal A}_l(\cdot,\tau, \sigma)\|_\infty&\leq&
(\tau-\sigma)Ce^{(\tau-\sigma)}
\tau^{-{1\over 2}}
\EN

\vs{3mm}

Given Lemmas 1 and 2, we may next solve (7) by the contraction
mapping principle. Thus, write (7) as
\be
\psi(\tau)=\psi^0(\tau)+{\cal N}(\psi,\tau)\equiv {\cal S}(\psi,\tau)
\en
where $\psi^0$ collects the linear and inhomogenous terms
that were bounded in Lemmas 1 and 2.

Consider now the following norm on
$C^0(\Bbb R)$. For $\psi\in C^0(\Bbb R)$, we set
\be
|\psi|_\tau=\tau^{2-\delta}\|(1+|\xi|^{3})^{-1}\chi\psi\|_\infty +
\tau^{{1\over 2} -\delta}\|(1-\chi)\psi\|_\infty , \en where
$\chi=\chi(\cdot,\tau)$. We have \be
C_1(\tau)\|\psi\|_\infty\leq|\psi|_\tau\leq C_2(\tau)\|\psi\|_\infty \en
for $C_1(\tau)>0$ and thus
$C^0(\Bbb R)$ is complete in the norm $|\cdot |_\tau)$.

Equation (20) is now solved for $\psi(\tau)\in{C^0(\bf R)}$ for $\tau\in
[\sigma,\sigma+\rho ]$,
with the norm
\be
\|\psi\|_\rho=\sup_{\tau\in[\sigma,\sigma+\rho]} |\psi(\tau)|_\tau.
\en
We shall choose below $\rho$ large enough and then take $\tau_0$ so that,
for $\sigma \geq
\tau_0$, we have, $\frac{\tau}{\sigma} \leq 1 + \frac{\rho}{\sigma} \leq
2$, $e^{c\rho} \leq
\tau_0^\delta$ and $A,A_l\leq
\tau_0^\delta$.
 Then, it is an
immediate corollary of Lemmas 1 and 2, that \be
\|\psi^0\|_\rho\leq C
\en
and we shall prove

\vs{3mm}

\no{\bf Lemma 3}. {\it ${\cal S}$ maps the ball
\be
B_0=\{\psi\in{C^0(\bf R)}\;|\:\|\psi-\psi^0\|_\rho\leq
\rho \tau^{-2\delta}\}
\en
into itself and, for $\psi_1$, $\psi_2\in B_0$,
\be
\|{\cal S}(\psi_1)-{\cal S}(\psi_2)\|_\rho\leq\lambda
\|\psi_1-\psi_2\|_\rho
\en
with $\lambda < 1$. Moreover, for $\psi \in B_0$, we can write
\be
{\cal N}(\psi,\tau)= \sum\limits_{m=0}^{2} \beta_mh_m +\beta^{\perp} +
\beta_l
\en
where, for $\tau \in [\sigma,\sigma + \rho]$,
\BE
|\beta_m(\tau)|&\leq& (\tau-\sigma)\tau^{-2}\;\;\;m=0,1\\
|\beta_{2}(\tau)|&\leq& (\tau-\sigma)\tau^{-3}\\
|\beta^\perp(\xi,\tau)|&\leq&(\tau-\sigma)(1+|\xi|^{3}) \tau^{-2}\\
\|\beta_l(\cdot,\tau)\|_\infty&\leq& (\tau-\sigma)\tau^{-{1\over 2}}
\EN}

\vs{3mm}

\no{\bf Remark}. Using the Lemmas, it is straightforward to show that (7)
has a $C^0$ solution
$\psi$. Using integration by parts and the regularity of the kernel $K
(\tau, \sigma)$ (see (41, 44)
below), one can show that this solution is actually smooth and is the
unique classical solution of equation (2.41).

\vs{3mm}

With Lemmas 1--3 we may now prove the Proposition. First,
writing $\psi=\psi^0+\psi^1$, we have the bounds (28-31) for
 $\psi^1$ and thus, combining these with (10)-(13), (16)-(19), we get
the following estimates for
the flow, for $\tau \leq \sigma+ \rho, \sigma \geq \tau_0$:
\BE
 & & |\psi_m(\tau)-e^{(1-{m\over 2})(\tau-\sigma)} \psi_m(\sigma)| \leq
(\tau-\sigma)Ce^{(\tau-\sigma)}\tau^{-2} \;\;\;m=0,1  \\
 & & |\psi_{2}(\tau)-({\sigma\over\tau})^2\psi_{2}(\sigma)| \leq
(\tau-\sigma)CA\tau^{-3}  \\
 & & |\psi^\perp(\xi,\tau)| \leq C(e^{-{1\over
2}(\tau-\sigma)} A+ e^{-(\tau-\sigma)^2}A_l +(\tau-\sigma))(1+|\xi|^{3})
\tau^{-2}  \\
 & & \|\psi_l(\cdot,\tau)\|_\infty
\leq C(A_l e^{-{(\tau-\sigma)\over p}}+A
e^{(\tau-\sigma)}+(\tau-\sigma)e^{(\tau-\sigma)})
\tau^{-{1\over 2}}
\EN

Now, we use (32-35) to prove the Proposition inductively. First, we prove
 the bounds (4-6) for
all times of the form $\tau_n=\tau_0+n \rho$, $n \geq 0$, with some
constants
$\tilde A, \tilde
A_l$. Then, it is easy to get (4-6) from (32-35) with
$\sigma=\tau_n$, with possibly other
constants, depending only on $\rho$, for all
times (for $m=2$, one uses inequality (38) below).

Next, we observe that, for $n=0, \tau=\tau_0, \psi$ is given by (1) and
$\|g\|_\infty \leq
\tau_0^{-2}$. We have
\BE
|\psi_0(\tau_0)+ \frac{a}{\tau_0} - d_0\gamma_0|+
|\psi_1(\tau_0)-d_1\gamma_1\tau_0^{-1/2}|\leq
C\tau_0^{-2}.
\EN
for nonzero constants $\gamma_0, \gamma_1$.
 From (1) and (36), it is easy to see that (4)-(6) hold for
 $n=0, \tau=\tau_0$ for a suitable
choice of $d_0,d_1$. Actually, one also sees from (36) that we may,
 instead
of varying $d_m$ vary $\psi_m(\tau_0)$. Let $\psi_m(\tau_0)$ be in the
interval $[-\tilde A\tau_0^{-2},\tilde A\tau_0^{-2}]$
for ${\tilde A}$ large.
 Let us assume now that we can find
$\psi_m(\tau_0)$ in that interval,
such that (4) holds for $m=0,1$, for all times,
with $A$ replaced
by
$\tilde A$.
Then, (5,6) hold,
using
(34,35): choose $\rho$ large enough so that
\BE
C(\tilde A e^{-\rho/2} + \tilde A_l e^{-\rho^2}+\rho) \leq \tilde A
\nonumber\\
C(\tilde A_l e^{-\rho/p} + \tilde A e^{ \rho} +\rho e^{\rho}) \leq \tilde
A_l
\EN
This is possible, for suitable $\tilde A, \tilde A_l$,
if we take $C \tilde A e^{ \rho} \leq \frac{\tilde A_l}{2}$ and
$\tilde A_l e^{-\rho^2}\leq \frac{\tilde A}{2}$, i.e. $Ce^{
\rho-\rho^2} \leq \frac{1}{4}$.

For (4) with $m=2$, we have
\be
|\psi_{2}(\tau)| \leq(\sigma\tau^{-1})^\delta \tau^{-2+\delta}+
(\tau-\sigma)CA\tau^{-3}\leq\tau^{-2+\delta}
\en
for $\tau \geq \sigma$, large enough.

It thus remains to show that
there exist $\psi_m(\tau_0)\in
[-\tilde A\tau_0^{-2},\tilde A\tau_0^{-2}]$, $m=0,1$, such
that (4) holds for all $\tau$. Suppose such
$\psi_m(\tau_0)$ did not exist. Set
$y=\tilde A^{-1}\tau_0^2(\psi_0(\tau_0), \psi_{1}(\tau_0))\in
{\cal C}=[-1,1]^{2}\subseteq{\Bbb R}^{2}$ and $\phi=\tilde
A^{-1}\tau^2(\psi_0(\tau)
,\psi_{1}(\tau))$. We have shown that $\phi=\phi(\tau,y)$ is continuous in
$\tau$ and $y$. Moreover, by the above assumption, for all $y$ there exists
a
first time $\tau(y)$, such that $\phi(\tau(y),y)\in\partial{\cal C}$. Also,
by
(32), the flow $\phi(\tau,y)$ is transversal to $\partial{\cal C}$ (by
induction, (32) holds up to time $\tau(y)$). This implies that $\tau(y)$
is continuous. Thus, $y\rightarrow\phi(\tau(y),y)$ is a continuous map from
the unit square ${\cal
C}$ in ${\Bbb R}^{2}$ to its boundary $\partial{\cal C}$, which is the
identity on the
boundary. Such a map can not exist, since ${\cal
C}$  is contractible to a point and this map would then provide a homotopy
between
the identity map $S^{1}\rightarrow S^{1}$ and the constant map. Thus we
can choose the $d_0,d_1$ such that (4), and hence all the other claims of
the
Proposition hold. \hfill$\Box$

To summarize, the logic in the choice of constants is as follows: first,
take $K$ in (2.57) large
enough, and $\delta$ small enough, so that various estimates hold.
For example, we shall use often the bound (see (2.7)):
 for $K$ in (2.57) large enough
\be
\int P(\xi) (1-\chi(\xi,\tau)) d \mu(\xi) \leq C(P)e^{-\tau}
\en
for any polynomial $P$, where $C(P)$ depends on $P$.
This
choice of $K$
and $\delta$ fixes the
constants appearing in the bounds used in the proof. Then, we take $\rho$
large enough so that
(34,35) iterate (see (36, 37))  .  Finally, take $\tau_0$ large, given
$\rho$ and the various constants
appearing in the proofs, so that we can write e.g. $\tau^{- \delta} \leq
e^{- c\rho}$ or $C \leq
\tau^\delta$ for $\tau \geq \tau_0$. In several
estimates below, we replace $\sigma$ by $\tau$,
which will be legitimate, using $\frac{\tau}{\sigma} \leq 2$.

\vs{5mm}

\no We will now prove Lemmas 1-3.

\vs{3mm}

\no{\bf Proof of Lemma 1}.
Let us denote $\tau-\sigma$ by $t$ and
$K(\tau,\sigma)$ by $K_t$. $K_t$ is
the fundamental solution of the linear equation
$\dot{K}=({\cal L}+V)K$ and we will use a
Feynman-Kac representation for it.
Since ${\cal L}$ is conjugated to the harmonic oscillator:
\be
e^{-\xi^2/8}{\cal L}e^{\xi^2/8} = \partial^2 - {\xi^2\over 16^2} +
{1\over 4}
+ 1 \en
we may write
\be
K_{t}(\xi,{\xi'}) = e^{t {\cal L}}
(\xi,{\xi'}) \int d\mu^{t}_{\xi{\xi'}}
(\omega)e^{\int\limits_0^{t} V(\omega(s),\sigma+s)ds}
\en
where $d\mu^{t}_{\xi{\xi'}}(\omega)$ is the oscillator
 measure on the continuous paths
$\omega:[0,t] \to {\Bbb R}$ with $\omega (0) = {\xi'},\
\omega (t) = \xi$, i.e. the Gaussian probability
measure with covariance kernel
\be C(s,s') =
\omega_0(s)\omega_0(s') + 2(e^{-{1\over 2}|s-s'|} -
 e^{-{1\over 2}|s+s'|} + e^{-{1\over 2}|2t - s'
+ s|} - e^{-{1\over 2}|2t-s'-s|}), \en
and mean $\int d\mu^{t}_{\xi{\xi'}}
(\omega) \omega(s)=\omega_0(s)$, where
\be
\omega_0(s) = ({\sinh} {t\over 2})^{-1}(\xi\ {\sinh} {s\over 2} +
{\xi'}\
{\sinh}{t - s\over 2}). \en
The kernel of $e^{t{\cal L}}$ is given
explicitely by Mehler's formula \cite{Si}
\be
e^{t {\cal L}}(\xi,{\xi'}) = [4\pi (1 - e^{-t})]^{-1/2}e^{t} \exp\
[-{(\xi e^{-t/2} - {\xi'})^2\over 4(1 - e^{-t})}]. \en

\vs{3mm}
Although the proof of Lemma 1 is long, most of it can
be understood easily by considering
(41), (44). If we replace $K_{t}$ by $ e^{t {\cal L}}$,
we understand the LHS of (10), coming from (2.11).
But the potential $V$, see (2.42), is
of order $\tau^{-1}$ for $\xi$ of order one. The precise estimate
 is done in Lemma 5 below and gives as a correction the RHS of (10).
In (11), the term on the LHS comes from $V$, as we saw in (2.52),
and the RHS is as in (10). For (12), the first term
on the RHS comes from the fact that $e^{t {\cal L}}$ contracts
$\theta^\perp$, which follows also from (2.11); however,
we shall use an integration by parts and the explicit formula (44),
in order not to expand as in (2.54). The second term in
the RHS of (12) is the contribution
to small $\xi$, coming from large $\xi'$. Looking at (44),
 we see that this contribution
is small for large $t$. Finally, in (13), the first
term in the RHS, i.e.
the contribution from large $\xi'$, is suppressed
because the potential is no longer small, while the
one coming from small $\xi'$
is controlled because $\psi_s$ is bounded by ${\cal O}(\tau^{-1/2})$.

\vs{3mm}

Let us now bound each term in (8).
Consider first $\theta_m$: let $k_m = h_m\Vert h_m\Vert^{-2}$. Then $
\theta_m (\tau) = (k_m,\chi_\tau K_t \psi (\sigma))$, where $\chi_\tau =
\chi (\xi,\tau)$, and $\psi_m(\sigma)=(k_m,\chi_{\sigma} \psi (\sigma))$.
We
write for $m=0,1$,
\BE
\chi_\tau K_t - e^{(1-\frac{m}{2})t} \chi_{\sigma}
= \chi_\tau (e^{t {\cal L}} - e^{(1-\frac{m}{2})t}) + \chi_\tau
(K_t-e^{t {\cal L}}) + e^{(1-\frac{m}{2})t} (\chi_\tau -
\chi_{\sigma}).
\EN
Consider the first term in (45).
Using (2,3) and (2.11), we have, writing $\chi=\chi_\tau$, $\psi =\psi
(\sigma)$,
\BE
(k_m,\chi (e^{t {\cal L}}-e^{(1-\frac{m}{2})t})\psi)& =&
\sum\limits_{r=0}^{2} [(e^{(1-{r\over
2})t}-e^{(1-\frac{m}{2})t})
\psi_r(k_m,\chi h_r)]\nonumber\\
&+ & (k_m,\chi (e^{t {\cal L}}-e^{(1-\frac{m}{2})t}) \psi^{\perp}) +
(k_m,\chi (e^{t {\cal L}}-e^{(1-\frac{m}{2})t})\psi_l).
\EN
For the first term in (46),
use (39) to get
\be
|(k_m,\chi h_r) - \delta_{mr}| \leq Ce^{-\tau}.
\en
\medskip
Indeed, by definition, $(k_m,h_r)=\delta_{mr}$.

For the two other terms and for later
purposes,  we need the following property of the kernel of  $e^{t {\cal
L}}$,
that follows  easily from the explicit expression (44):

\vs{3mm}

\no{\bf Lemma 4.}
{\it Let  $|\psi ({\xi'})| \leq (1 + |{\xi'}|)^p$, for $p \geq 0$.  Then}
\be
|(e^{t {\cal L}}\psi)(\xi)| \leq C e^{t}(1 + e^{-t/2}|\xi|)^p
\en

\vs{3mm}

\no
With Lemma 4 and equations (5, 39), we get, since $(k_m,(e^{t {\cal L}} -
e^{(1-\frac{m}{2})t}) \psi^\perp)=0$, for $m \leq 2$,
\be
|(k_m,\chi (e^{t {\cal L}} -
e^{(1-\frac{m}{2})t}) \psi^{\perp})| = |(k_m,(1 - \chi)((e^{t {\cal
L}}-1) -
(e^{(1-\frac{m}{2})t}-1))
\psi^{\perp})| \leq C A te^t e^{-\tau}
\en
Indeed, we may write $e^{t {\cal
L}} -1=\int_0^t ds {\cal L} e^{s {\cal L}}$,
and use the fact that $(1-\chi)k_m \in {\cal D}({\cal L}) $
and ${\cal L}(1-\chi)k_m  $ has support in $|\xi| \geq K$,
which follows from the smoothness of $\chi$.

Finally, for the last term in (46), using (6), and reasoning as above,
\BE
& & |(k_m,\chi (e^{t {\cal L}} -
1)\psi_l)| \leq
 A_l \sigma^{-{1\over 2}}
\int d\mu(\xi) d\xi' |{\cal L}(k_m \chi_\tau)(\xi)|
\nonumber \\
&&\cdot\int_0^t ds  e^{s {\cal L}}(\xi,\xi')
(1 -
\chi(\xi',\sigma))
\leq   Ct e^{-\tau}.
\EN
and a similar bound for the term with
$1-e^{(1-\frac{m}{2})t}$.
Indeed, if we insert  (44) and  (2.7) into (50), we end up
with the estimate
\be
\sup_{\xi\leq 2K\tau^{1/2},\xi'\geq K\sigma^{1/2}}
e^{-\frac{\xi^2}{4}-{(\xi e^{-s/2} - {\xi'})^2\over 4(1 - e^{-s})}}
\leq e^{-2\tau}.
\en
for $K$ large enough.
We may use the square root in the LHS of (51) to control the integrals in
(50) and the factor $e^t$ in (48).
 The constant
$A_l$ in (50) is bounded by the factor $\sigma^{- \frac{1}{2}}$, for
$\sigma$
large enough.

\vs{3mm}

 For the second term in (45),  we write again
\be
(k_m,\chi (K_{t}
- e^{t {\cal L}})\psi) = \sum\limits_{r=0}^{2} (k_m,\chi (K_{t} -
e^{t
{\cal L}})h_r)\psi_r + (k_m,\chi (K_{t} - e^{t {\cal
L}})(\psi^{\perp} +
\psi_l)).
\en
Now we need some properties of $K_t$:

\vs{3mm}

{\bf Lemma 5.} {\it The kernel  $K_t(\xi,{\xi'})$  given by}
 (41){\it satisfies
\be
K_t(\xi,{\xi'}) = e^{t {\cal L}}(\xi,{\xi'})(1 + {1\over
\tau}P_{2}(\xi,{\xi'}) + R(\xi,{\xi'}))
\en
where  $P_{2}$  is a polynomial
\be
P_{2}(\xi,{\xi'}) = \sum\limits_{m+n\leq 2} p_{mn}\xi^m{\xi'}^n
\en
with  $|p_{mn}| < C t$, and $R$  is bounded by
\be
|R(\xi,{\xi'})| \leq C t (1+t)\tau^{-2}(1+|\xi| + |{\xi'}|)^{4}
\en
Moreover,}
\be
|(k_{2},(K_t-({\sigma\over\tau})^2)h_{2})|\leq Ct(1+t) \tau^{-2}
\en

\vs{3mm}

\no Using (53)--(55),(4), Lemma 4 and (2.7), we have, for $m\leq 2$,
\BE
& & |(k_m,\chi (K_{t} - e^{t {\cal L}})h_r)\psi_r| \leq
CAt \tau^{-3} \hspace*{10mm} r=0,1 \nonumber \\
& & |(k_m,\chi (K_{t} - e^{t {\cal L}})h_2)\psi_2| \leq
C t \tau^{-3+\delta}.
\EN
By (5) and Lemma 4, the $\psi^{\perp}$--term in (52) also satisfies a bound
like the first inequality in (57) and the  $\psi_l$-term  is bounded as in
(50).  For the last term in
(45), involving $\chi_\tau-\chi_{\sigma}$, we can bound its contribution by
$C
t e^{-\tau}$, using (2.7) as in (39). Hence, combining
(45,47,49,50,57), we
get
\be
|\theta_m - e^{ (1-{m\over 2})t}\psi_m| \leq C t
\tau^{-3+\delta},\hspace*{10mm} m=0,1
\en
For $m=2$, we write first
\BE
\chi_\tau K_t - \left(\frac{\sigma}{\tau}\right)^2\chi_{\sigma}
= \chi_\tau \left(K_t - \left(\frac{\sigma}{\tau}\right)^2\right) +
(\chi_\tau - \chi_{\sigma}) \left( \frac{\sigma}{\tau} \right)^2
\non
\EN
Then, combine the previous bounds with (56), using only the first
inequality in
(57) since we use (56) for the $r=2$ term. We get:
\be
|\theta_{2} - ({\sigma\over\tau})^2\psi_{2}|  \leq C A t \tau^{-3}.
\en
This proves (10,11).

Next, consider  $\theta^{\perp}$  in (8).  Let
$P^{\perp}$  be the projection in  $L^2({\Bbb R},d\mu)$  on the
corresponding subspace.  We write, using (2,3),
\be
\theta^{\perp} = P^{\perp}\chi K_{t}\psi = P^{\perp}\chi K_{t}
\psi^{\perp}  + \sum^2_{r=0} \psi_rP^{\perp}\chi K_{t}h_r +
P^{\perp}\chi K_{t}\psi_l
\en
and consider again the various terms separately.

For the first term, we can write
\be
(K_{t}\psi^{\perp}) = \int d\xi'M(\cdot,\xi')f(\xi')
\en
where,
\be
M(\xi,\xi') = e^{{\xi'}^2/4}K_{t}(\xi,\xi'),\ \
f(\xi') = e^{{-{\xi'}^2}/4}\psi^{\perp}(\xi',\sigma)
\en
i.e., see (41,44),
\BE
M(\xi,\xi')& =& [4\pi (1 - e^{-t})]^{-1/2}e^{t}e^{\xi^2/4}\
e^{-{(\xi - e^{-t/2}\xi')^2\over 4(1 - e^{-t})}}
\langle e^V\rangle (\xi,\xi')\non\\
&\equiv &N(\xi,\xi')\langle e^V\rangle (\xi,\xi')
\EN
where we used the identity
\BE
\frac{(\xi e^{-t/2}-\xi')^2}{1-e^{-t}}-{\xi'}^2 = - \xi^2+
\frac{(\xi-e^{-t/2}\xi')^2}{1-e^{-t}}
\EN
and the notation
\BE
\langle e^V\rangle(\xi,\xi') = \int d\mu^t_{\xi{\xi'}}(\omega)
e^{\int\limits_0^{t} V(\omega (s),\sigma+s)ds} \nonumber
\EN
Now, $(\psi^{\perp},h_m) = 0,\ m \leq 2$  means (see (2.7))
$\int f({\xi'}){\xi'}^md{\xi'} = 0,\ m \leq 2$.
Thus, let  $f^{(-m)}$  be the $m$:th antiderivative of  $f$, i.e.
\be
f^{(-m-1)}(\xi) = \int\limits_{-\infty}^\xi d{\xi'}f^{(-m)}({\xi'}).
\en
We have

\vs{3mm}

{\bf Lemma 6.} {\it For $f$ defined in
}(62),{\it
\be
|f^{(-m)}(\xi)| \leq CA\tau^{-2}(1 + |\xi|)^{3-m}e^{-\xi^2/4}
\en
for}  $m \leq 3$.

\vs{3mm}

Now write (61) by integrating by parts,
\BE
(K_{t}\psi^{\perp})(\xi)& =& \sum\limits_{r=0}^{2} \int
\partial_{\xi'}^r N(\xi,{\xi'})\partial_{\xi'}\langle
e^V\rangle (\xi,{\xi'})f^{(-r-1)}({\xi'})d{\xi'}
\non\\&&
+ \int \partial_{\xi'}^{3} N(\xi,{\xi'})\langle
e^V\rangle (\xi,{\xi'})f^{(-3)}(\xi')d{\xi'}.
\EN
We need the integration by parts formula for Gaussian measures \cite{GJ}:
\BE
\partial_{\xi'}\langle e^V\rangle(\xi,\xi')& =& {1\over 2} \int_0^{t}
dsds'\partial_{\xi'}
C(s,s')
\int d\mu^{t}_{\xi{\xi'}}(\omega)V'(\omega(s),\sigma+s)
V'(\omega(s'),\sigma+s')
e^{\int V}\non\\&&
+{1\over 2} \int_0^{t} ds\partial_{\xi'}
C(s,s)
\int d\mu^{t}_{\xi{\xi'}}(\omega)V''(\omega(s),\sigma+s)
e^{\int V}
\EN
Since (see (2.42))
\BE
V &\leq &\frac{C}{\tau}\non\\
|{d^n V\over d\xi^n}| &\leq & C\tau^{-n/2},\; n = 0,1,2
\EN
and $C(s,s')$ is given by (42), we have
$\int d\mu^{t}_{\xi{\xi'}}(\omega)
e^{\int V}\leq C
$ and
\be
|\partial_{\xi'}\langle e^V\rangle (\xi,{\xi'})| \leq
C\tau^{-1}t(1+t)(|\xi| + |{\xi'}|).
\en
As for  $\partial^r_{\xi'}N$, we get from (63), for  $t > 1$,
\be
|\partial^r_{\xi'}N(\xi,{\xi'})|
 \leq Ce^{-{rt\over 2}}(|\xi| + |{\xi'}|)^re^{\xi'^2/4}
e^{t {\cal L}}(\xi,{\xi'}).
\en
where we used (64,44) to rewrite the RHS.
Thus, from (66), (67), (70), (71) and Lemma 4, using
$\tau^{-1}t(1+t)
\leq e^{-3t \over 2}$ for $\tau$ large, we get,
\be
|(K_{t}\psi^{\perp})(\xi)| \leq CA\tau^{-2}e^{-t/2} (1 + |\xi|)^{3}.
\en
To control $P^{\perp}\chi K_{t}\psi^{\perp}$,
we use the following remark; let $X(\xi)$
satisfy
$$
|X(\xi)|\leq \eta (1 + |\xi|)^{3}
$$
Then, using (2.7), we have
$$
|(k_m,X)| \leq C\eta.
$$
Hence, $P^{\perp}X(\xi)  = X(\xi) -
\sum\limits_{m=0}^{2} (k_m,X)h_m(\xi)$
satisfies
\be
|P^{\perp}X(\xi)|\leq C\eta (1 + |\xi|)^{3}
\en
So,  $P^{\perp} \chi K_{t}\psi^{\perp}$  satisfies a
bound like (72).  If  $t \leq 1$, since the derivatives in (71)
bring extra factors of $t^{-1}$, so we do not integrate by parts as in
(67), but
derive the bound (12) for that term in (60) directly from Lemma 5, Lemma 4
and
(5).

Now consider the second term in (60).  Since  $K_{t}$  is given by
Lemma 5, we obtain
\be
|\psi_r(\chi K_{t}h_r)(\xi) -\psi_r e^{t(1-{r\over 2})}
(\chi h_r)(\xi)| \leq
CA\tau^{-3+\delta+1/2} e^{t}(1 + |\xi|^3).
\en
Indeed, we get $A\tau^{-2+ \delta}$ from (4); for the $P_2$ term in (53),
we have
$\tau^{-1}$ and, using Lemma 4,
$$
|e^{t {\cal L}} (P_2 h_r) (\xi) | \leq Ce^{t}(1+|\xi|^4)
$$
since $r \leq 2$.
But, on the support of $\chi, |\xi| \leq 2K \tau^{1/2}$, so we can replace
one
power of $|\xi|$ by $2K\tau^{1/2}$. Similarly, for $R$ in (53), we get from
(55)
and Lemma 4, a bound with $\tau^{-2}$ and $(1+ |\xi|)^6$ and we control
$|\xi|^3$ by $\tau^{3/2}$. Now, using (73), we get a bound
like (74) on $P^\perp
(\psi_r(\chi K_{t}h_r)(\xi) -\psi_r e^{t(1-{r\over 2})}
(\chi h_r)(\xi)) $.
We still have to consider
$\psi_r e^{t(1-{r\over 2})}P^\perp \chi h_r, r \leq 2$.
But, by definition,
$P^\perp h_r=0$, and we can replace $\chi$ by $(1-\chi)$, and use
$$
|(1-\chi)h_r| \leq \tau^{- 1/2} (1+ |\xi|)^3
$$
since $r \leq 2$ and $|\xi| \geq K \tau^{1/2}$ on the support of
$(1-\chi)$.
Then, by (73) again, $P^\perp (1-\chi)h_r$ satisfies a similar bound.
Hence, the
second term of (60) has a bound like the RHS of (74).

For the last term in (60), we use the bound (69) on $V$,
to get
 From (41), $K_t (\xi,\xi') \leq C e^{t {\cal L}}
(\xi,\xi')$.
Then, using (6) and (44), we have
\BE
&&\|(1+|\xi|^{3})^{-1}\chi
K_t\psi_l\|_\infty\leq  CA_l e^{ t}\tau^{-1/2}
\sup_{\xi,\xi'}(1+|\xi|^3)^{-1}\non\\
&&\cdot e^{-c(\xi
e^{-t/2}-\xi')^2}\chi(\xi,\sigma+t) (1-\chi(\xi',\sigma))\leq
\left\{ \begin{array}{ll}
CA_l\tau^{-2}& t\leq t_0\\
e^{-\tau}& t > t_0
\end{array}\right.
\EN
for a suitable $t_0$. Indeed, for $t$ large,
we get $e^{-2\tau}\leq e^{-\tau}(CA_l e^{ t})^{-1}$ from
$e^{-c(\xi e^{-t/2}-\xi')^2}$ and the characteristic functions, while,
for
$t$ small, we get $\xi \simeq \xi'$ and therefore $(1+|\xi|^3)^{-1} \leq
c
\tau^{-3/2}$ from $(1-\chi(\xi',\sigma))$. Then, proceeding as for (72,
73),
we get a bound on the last term of (60), which can be written as $C A_l
e^{-t^2}\tau^{-2}(1+|\xi|^3)$ for $\tau$
large. This bound is of course rather arbitrary, but convenient.

Hence, combining (72),(74) and
(75), and using $\tau^{- 1/2 +\delta}e^{t}\leq e^{-t/2}$ in (74), we have
\be
|\theta^\perp(\xi,\tau)|\leq C \tau^{-2} (A e^{-t/2} + A_l e^{- t^2})
(1+|\xi|^3).
\en
This proves (12).

Only $\theta_l$ remains to be bounded. We have (see (9)),
\be
\theta_l=(1-\chi)K_t\psi=(1-\chi)K_t(\psi_s+\psi_l).
\en
By (3, 4, 5), we have, using $\chi |\xi| \leq 2 K \sigma^{1/2}$, $|\psi_s|
\leq
C A
\sigma^{-1/2}$. Now, use $K_t \leq C e^{t {\cal L}}$ (from
(69) and (41)) and
(48) with $p=0$, to get
\be
\| K_t \psi_s \| \leq C A e^{ t} \tau^{- \frac{1}{2}}.
\en
This gives a bound on the first term of (77).

For the second term, we use

\vs{3mm}

\no{\bf Lemma 7}.  {\it Let  $\chi$  be the  function} (2.57).
{\it Then,}
\be
\Vert K_t(1 - \chi)\Vert_{\infty} \leq C e^{-t/p}.
\en

\vs{3mm}

\no Thus,
\be
\|K_t\psi_l\|_\infty\leq CA_l e^{-t/p}\tau^{-1/2}
\en
The bound (13) for $\theta_l$ follows from (78) and (80).
\hfill $\Box$

Now, we shall prove Lemmas 5-7 that were used in the proof of Lemma 1.

\vs{3mm}

\no{\bf Proof of Lemma 5}. We start with the Feynman-Kac formula (41). Let
\be
M(\lambda) = \int d\mu^{t}_{\xi\xi'}(\omega)
e^{\lambda \int\limits_0^{t} V(\omega (s),\sigma+s)ds}.
\en

$M$  is  $C^2$  in  $\lambda$  (in fact  $C^{\infty}$) and
\be
M(1) = 1 + \int\limits_0^{t} ds\int d\mu^{t}_{\xi\xi'}(\omega)
V(\omega(s),\sigma+s) + \tilde M
\en
with
\be
\tilde M = \int\limits_0^1 d\lambda (1-\lambda) \int^t_0 dsds' \int
d\mu^{t}_{\xi\xi'}(\omega)V(\omega (s),\sigma+s)
V(\omega (s'),\sigma+s')
e^{\lambda \int\limits_0^{t} V(\omega(u),\sigma+u)du}.
\en
We have the following bounds for $V$ given by (2.42):
\BE
V&\leq& \frac{C}{\tau}\\
|V(\omega,\tau)|&\leq& C \tau^{-1}(1+|\omega|)^{2}\\
V(\omega,\tau)&=&\tau^{-1}Q(\omega)+{\tilde V}(\omega,\tau)
\EN
with $Q(\omega)$  a polynomial of degree  $2$ in $\omega$ with bounded
coefficients  and where  $\tilde V$  satisfies
\be
|\tilde V(\omega,\tau)| \leq {C\over \tau^2}(1 + |\omega|)^{4}.
\en
Insert now (86) in (82) and, in (83), insert (84) in the exponent
and (85) elsewhere. Using formulas (42), (43) for
the covariance of $\mu$, the claims (53)-(55) follow.
$P$ is produced by $Q$
in (86), while $R$ collects ${\tilde M}$ and ${\tilde V}$.

Finally, we prove estimate (56). We want to show
that the contribution of the second term in (82) to $(k_{2},
K_t h_{2})$ is as claimed in (56).

Note that the term ${\tilde M}$ in (82) can be absorbed into the RHS
of (56). Also, by (86), we may replace $V$ in (82) by $\tau^{-1}
Q$, the error again being absorbed into the RHS of (56).

Using (14) and (2.42), we compute
\be
\tau^{-1}Q(\omega)=-{pb\over \tau(p-1)^2}h_{2}(\omega)=
\frac{-h_2(\omega)}{4
\tau}
\en

\vs{2mm}

\no We could now calculate the Gaussian integral in (82) directly,
using the covariance (42) and the fact that $Q$ is a polynomial.
The result, however, can be obtained directly, by noting that the above
estimates
imply that
\be
|\frac{d}{d \tau} (k_2,K_t h_2)-\frac{d}{d \tau}\psi_2^{(0)}(\tau)|\leq
C(1+t)\tau^{-2}
\en
(with $t=\tau-\sigma$)
where $\psi_{2}^{(0)}$ solves the equation
\be
\frac{d}{d \tau} \psi_{2}^{(0)}=(k_{2},\tau^{-1}Q(\cdot)h_{2})
{\psi}_{2}^{(0)}=-{(h^2_{2},h_{2})\over 4 \tau\|h_{2}\|^2}
{\psi}_{2}^{(0)}=-2\tau^{-1}{\psi}_{2}^{(0)}
\en
with initial condition $\psi^{(0)}_2(\sigma)=1$.
Indeed, we may first replace, with error bounded by (89),
$K_t$ by $e^{t{\cal L}}(1+\int_0^t ds
\frac{\langle Q(\omega(s))\rangle (\xi, \xi')}{\sigma +s})$ (use (82-87)).
Then, using
$e^{t{\cal L}}k_2=k_2$, we replace $(k_2,K_t h_2)$ by
$$
e^{(k_2(\cdot),\int_0^t ds \frac{\langle Q(\omega(s))\rangle
(\cdot, \cdot)}{\sigma +s}h_2(\cdot))},
$$
where, again, the error is bounded by (89).
 Differentiating gives (89, 90).
The solution of (90) is
${\psi}_{2}^{(0)}(\tau)= ({\sigma\over \tau})^2$, which
yields (56).\hfill $\Box$
\vs{3mm}

\no{\bf Proof of Lemma 6}. We show: let $\int fdx =0$ and
$|f(x)|\leq A(1+|x|^p)e^{-x^2/4}$ then
\be
|f^{(-1)}(x)|\leq CA(1+|x|^{p-1})e^{-x^2/4}.
\en
The claim then follows by induction in $m$ using the fact that $\int
f(x)x^mdx=0$ for $m \leq 2$ implies $\int f^{(-m)} (x) dx = 0$ for
$m\leq2$;
(91) follows from a simple calculation: let e.g. $x>0$. Then, since $\int
fdx=0$,
\BE
&&|f^{(-1)}(x)|\leq A\int_x^\infty(1+|y|^{p})e^{-y^2/4}dy\leq
\non\\&&
A\int_{x^2/4}^\infty u^{-{1\over 2}}(1+|u|^{p/2})e^{-u}du\leq
CA(1+|x|^{p-1})e^{-x^2/4}
\non
\EN
\hfill$\Box$
\vs{3mm}

\no{\bf Proof of Lemma 7.} We will use both the oscillator formula (41)
and another Feynman--Kac formula for  $K_{t}$,  in terms of the Wiener
measure, which follows from the conjugation (40):
\be
K_{t}(\xi,\xi') =
\exp\ [(1 + {1\over 4})t + {1\over 8}(\xi^2 - \xi'^2)]
\int d\nu^{t}_{\xi\xi'}(\omega)e^{U(\omega)}
\en
where
\be
U(\omega) = \int\limits_0^{t} (V(\omega (t),\sigma+t) - {1\over 16}
\omega (t)^2)dt
\en
and  $d\nu^{t}_{\xi\xi'}$  is the Wiener measure on continuous paths
$\omega : [0,t] \to {\Bbb R},\ \omega (0) = \xi',\ \omega (t) = \xi$
with the normalization
\be
\int d\nu^{t}_{\xi\xi'}(\omega) = e^{t \partial^2}(\xi,\xi').
\en
We want to estimate
\be
\int K_{t}(\xi,\xi')(1-\chi (\xi',\sigma))d\xi'
\leq Ce^{-t /p}
\en
uniformly in $\xi$.
Let us divide the integral into two regions:

\vs{2mm}

\no(a) Let
$|\xi e^{-t/2} - \xi'| \geq |\xi'|/4$.  By (84), (41)
and (44)
\be
K_{t}(\xi,\xi') \leq C (1-e^{-t})^{- \frac{1}{2}} e^{t}
e^{- \frac{c\xi'^2}{1-e^{- t}}}.
\en
and that contribution to (95) is bounded by $e^{-\sigma}$ for $K$ in (2.57)
(and $\sigma$) large enough.

\vs{2mm}

\no (b) Let $|\xi
e^{-t/2} - \xi'| \leq { 1\over  4}|\xi'|$, hence, for
$\xi' > 0$, say, $\xi \in [{ 3\over  4}
e^{t/2}\xi',{ 5\over  4}
e^{t/2}\xi']$.  we use the representation (92) and condition on  the
first
time  $t$ such that  $\omega (t) = { 1\over  2}\xi'$, if $\omega$ visits
$\frac{1}{2} \xi'$ at all. So,
\BE
\int d\nu^{t}_{\xi,\xi'}(\omega)e^{U(\omega)} &=&
\int\limits_0^{t}dt_1\int d\nu^{t_1}_{{1\over 2}\xi',\xi'}
(\omega_1 >{\xi'\over 2})
\int d\nu^{t -t_1}_{\xi,{1\over 2}\xi'}(\omega_2)
e^{U(\omega_1\cup \omega_2)} \non\\
&+&  \int d\nu^{t}_{\xi\xi'}(\omega)\chi
(\omega > {1\over 2}\xi')e^{U(\omega)}.
\EN
where $d\nu^t_{a,b} (\omega > a)$ is the measure on paths $\omega$ such
that
$\omega (0)=b,\omega (t)=a$, and $\omega(s) >a$, for $s<t$, defined by
$$
\int F(\omega) d\nu^t_{a,b} (\omega > a) = 2\frac{d}{d x} \int F(\omega) d
\nu^t_{x,b} (\omega) |_{x=a}.
$$
One can check that this defines an expectation and that (97) holds by the
method
of images. Below we shall only use the formula
$$
\int d\nu^t_{a,b} (\omega >a) =
\frac{(b-a)}{(4 \pi t^3)^{\frac{1}{2}}} \; e^{- \frac{(a-b)^2}{4t}}
$$
which is the probability density that $t$ is the first time at which
$\omega$,
starting from $b$, reaches $a$. Hence,
\be
\int^t_0 d t_1 \int d\nu^{t_1}_{\frac{1}{2} \xi',\xi'} (\omega_1 >
\frac{\xi'}{2}) \leq 1
\en
For the second term in (97)
(this is where the extra contraction in large  $\xi$
comes from!), since  $|\omega (t)| \geq { K\over  2}\tau^{1/2}$
(because of the characteristic function in (95)),
we have from (93, 2.42),
\BE
U(\omega)& \leq& [-{p\over p - 1} + p\left(\left(
\frac{c}{K}\right)^{\frac{1}{p-1}}+\frac{a}{\tau}\right)^{p-1}]t -
{1\over 16}
\int\limits_0^{t}\omega (t)^2dt\non\\
&\leq&  -(1+\frac{1}{p})t -{1\over 16} \int\limits_0^{t} \omega
(t)^2dt
\nonumber
\EN
for $K$ (and $\tau$) large enough. Using (92, 41), the contribution
of the second term in (97) to
(95) (see (92)) is then bounded by
\be
e^{-(1+\frac{1}{p})t}\int e^{t {\cal L}}(\xi,\xi')d\xi' \leq
e^{-t/p}
\en
since  ${\cal L}1 = 1$.
\medskip
For the first term in (97), use (84) to bound
\be
U(\omega_1\cup \omega_2) \leq -{1\over 16}\int\limits_0^{t -t_1}
\omega_2(t)^2dt + \frac{Ct}{\tau}
\en
so that, using (98), this term is bounded by
\be
C \sup_{t_1 \in [0,t]}
(e^{(t -t_1)(\partial^2-{\xi^2\over 16})}
(\xi,{\xi'\over 2}))\cdot.
\en
Hence, using (40,92) its contribution to  $K_{t}$  is bounded by
\be
C\sup_{t_1 \in [0,t]}e^{(t -t_1){\cal L}}
(\xi,{\xi'\over 2}).
\en
 From the Mehler formula (44), since (let  $\zeta = t - t_1$)
\be
(\xi e^{-\zeta/2} - {\xi'\over 2})^2 \geq ({\xi'\over 4})^2
\en
(recall that  $\xi \geq { 3\over  4}e^{t/2}\xi')$, this term, inserted
in
(95), contributes  ${\cal O}(e^{-c{\xi'}^2}) = {\cal O}(e^{-\sigma})$.
Thus the claim follows by
combining this bound with (96, 99).\hfill$\Box$

\vs{3mm}

\no{\bf Proof of Lemma 2}.
We shall prove the following bounds for $\alpha$ defined in (2.44):
\BE
|\alpha_m(s)|&\leq& Cs^{-2}\;\;\;m=0,1\\
|\alpha_{2}(s)|&\leq&Cs^{-3}\\
|\alpha^\perp(\xi,s)|&\leq& C(1+|\xi|^{3}) s^{-2}\\
\|\alpha_l(\cdot,s)\|_\infty&\leq& Cs^{-{1\over 2}}
\EN
Then, using Lemma 1 with $\psi(\sigma)$ replaced by $\alpha(s)$,
$K(\tau,\sigma)$
replaced by $K(\tau,s)$, and integrating over $s$, we get (16-19).
We shall first show that, when $a=2b(p-1)^{\frac{2p-1}{1-p}}$, (104) holds
and
then show that, if we choose $b=b^*$, (105) holds also. For (104), we note
that,
using (2.7)
\be
|(h_m,-{\dot\varphi}_b+W\eta_0-{\dot\eta}_0+M(\eta_0))|\leq
Cs^{-2}
\en
Indeed, we may Taylor-expand $M(\eta_0)$ in the scalar product.
 From the expression (2.34) for $\varphi_b$, we deduce (we
set $c_p=(p-1)^{1\over 1-p}$)
\be
|(h_m,\varphi_b''+c_p{2b\over s (p-1)^2})|\leq Cs^{-2}
\en
but $\frac{c_p}{(p-1)^2} = (p-1)^{\frac{2p-1}{1-p}}$ and therefore, the
contribution of $\varphi''_b$ to order $s^{-1}$ and of $\eta_0 =
\frac{a}{s}$ in (2.44) cancel each other (actually, in (109) only $m=0$
is
needed, since $h_1$ is orthogonal to constants). This proves (104).

Next, consider (105).
Since $h_{2}$ is orthogonal to constants,
$({\eta}_0,h_{2})=({\dot\eta}_0,h_{2})=0$, and
we want to show that
\be
|(h_{2},\varphi_b''-{\dot\varphi}_b+W\eta_0+M(\eta_0))|\leq Cs^{-3}
\en
Again we Taylor-expand and get
\BE
&&|(h_{2},\varphi_b''-
{\dot\varphi}_b-c_p({pb^2\over 2(p-1)^4}\p^2\xi^{4}
-{b\over(p-1)^2}\xi^{2})s^{-2})|\leq Cs^{-3}\\
&&|(h_{2},(W+{pb \xi^2\over (p-1)^2 s})\eta_0)|
\leq Cs^{-3}\\
&&|(h_{2},M(\eta_0))|\leq Cs^{-3}
\EN
because the only term in $s^{-2}$ coming from $M (\eta_0)$ is constant,
i.e.
is orthogonal to $h_2$. Now, we compute $(h_2,\xi^2) = \| h_2
\|^2=8,(h_2,\partial^2\xi^4)=96$. Thus, all terms of order
$s^{-2}$ will cancel if the following equation holds:
$$
c_p \left(\frac{96 pb^2}{2(p-1)^4} - \frac{8b}{(p-1)^2} \right) =
\frac{8pba}{(p-1)^2}
$$
Using our choice of a, this is an equation for b, whose solution is $b^* =
\frac{(p-1)^2}{4p}$. So, (105) holds.

The bounds (106,107) are rather trivial. Since $\alpha$ is a smooth
function
of $\xi
s^{-1/2}$, with bounded derivatives, we would get a bound with
$s^{-3/2}$
in (106) from a Taylor expansion. But we have an extra $s^{-1}$ factor
coming
either from $\eta_0$ or from derivatives; (107) is proven by inspection.
\hfill$\Box$

\vs{3mm}

\no{\bf Proof of Lemma 3}. From (1.11), (2.5) together with
(2.43), we get
\be
|N(\psi(\xi,s))|\leq C(|\psi(\xi,s)|^{\tilde p} + e^{-cs}).
\en
where ${\tilde p}=\min(p,2) >1$.
Equations (21), (24), (25) and
\be
s^{-2+\delta}(1+|\xi|^3)\chi_s \leq C \tau^{- \frac{1}{2}+\delta}
\en
which holds for $\sigma \leq  s$, $\tau \leq \sigma +\rho$,
imply, for $\psi \in B_0$,
\be
\|\psi(\cdot,s)\|_\infty\leq C \tau^{-1/2+\delta}.
\en
So, we have, using $K_{\tau -s}(\xi,\xi')
\leq C e^{(\tau -s)
{\cal L}}(\xi,\xi')$
(from (41) and (84)) and ${\cal L}1=1$ for the second term in
(114),
\be
|{\cal N}(\psi)(\xi,\tau)|\leq C((\tau^{-{1\over
2}+\delta})^{{\tilde p}-1}\int_{\sigma}^\tau
ds\int d\xi'K_{\tau -s}(\xi,\xi')|\psi(\xi',s)|+(\tau -\sigma) e^{(\tau
-\sigma)}
e^{-c\tau}).
\en
We want to show that
\be
|K_{\tau -s} |\psi||_\tau\leq C e^{(\tau -s) } |\psi|_s
\en
Write $ |\psi | = | \psi \chi_s | + | \psi (1- \chi_s)|$ and estimate in
(118)
separately the large
$\xi$ and the small $\xi$ parts of the norm (21). We have, using $K_{\tau
-s}(\xi, \xi')
\leq C e^{(\tau -s)
{\cal L}}(\xi, \xi')$ (coming from (84)), and (48),
$$
| K_{\tau -s} | \psi \chi_s || \leq Ce^{ (\tau -s)} \tau^{-2+ \delta}
(1+ | \xi
|^3) | \psi |_s
$$
while, for $\chi_\tau K_{\tau -s} | \psi (1- \chi_s)|$, we can use a bound
like (75). This
proves (118) for $| \xi
| \leq 2 K \tau^{1/2}$. For large $| \xi |$, we can use (78,80) and the
bound (116).

Now, for $\delta$ small and $\tau$ large
(so that $C(\tau^{-{1\over
2}+\delta})^{{\tilde p}-1}e^{\rho}\leq \tau^{-2\delta}$), we get from
(117,118), and (23-25),
\be
\|{\cal S}(\psi)-\psi^{(0)}\|_\rho =\|{\cal
N}(\psi)\|_\rho \leq \rho\tau^{-2\delta}
\en
as required. The proof of (26) is similar.

To prove of (28, 29), write $N =\chi_sN + (1-\chi_s) N$, and use
instead of (114)
$$
|\chi_sN(\psi(\xi,s))|\leq C(\chi_s|\psi(\xi,s)|^{2} + e^{-cs}).
$$
because, for $|\xi|$ small, we can Taylor-expand $N$.
By (21,23-25), we have $|\chi \psi|^2 \leq C\tau^{-4+2 \delta} (1+ | \xi
|^6)$. Now,
 using $K_{\tau -s}
(\xi, \xi')
\leq C e^{(\tau -s)
{\cal L}}(\xi, \xi')$, Lemmas 4 and (2.7), we get (28,29) for $\chi_sN$.
For $(1-\chi_s)N$, we use the bound (114) on $N$ and (39,51). The
proof of (30,31) follows immediately from (119) and the definition of the
norm (where we can of course replace $\rho$ by $\tau-\sigma$). \hfill$\Box$

\section{ The proof, $k>1$}

\setcounter{equation}{0}

We need to study the equation (2.2) with $F=0$:
\be
\dot{\varphi} = L^{-2}_{\tau}\varphi'' - {1\over 2k}\xi
\varphi' - {1\over p-1}\varphi + \varphi^p
\en
and with initial data as in (1.14):
\be
\varphi (\xi,\tau_0) = f^*_b(\xi)(1 + \sum\limits_{i=0}^{2k-1} d_i\xi^i
(p-1 + b\xi^{2k})^{-1}) +
g(\xi)
\en
We want to prove the

\vs{3mm}

\no{\bf Theorem 4}. {\it
There exist  ${\bar \tau} < \infty$  and  $\varepsilon > 0$ such that\ for
$\tau_0 > {\bar \tau}$  and  $g$  in  $C^0({\Bbb R})$  with
$
\Vert g\Vert_{\infty} < \varepsilon$
there are constants  $d_i \in {\Bbb R}$  such that\ the equation}
(1){\it with the
initial data} (2) {\it has a unique classical solution, which satisfies
\be
\Vert \varphi (\cdot,\tau) - f^*_{b^*}(\cdot)\Vert_{\infty} \to 0\en
as $\tau \to \infty$,
for some  $b^* > 0$, where  $b^* \to b$  as  $\varepsilon \to 0$  and
$\tau_0 \to \infty$.}

\vs{3mm}

We reduce the proof of the Theorem again to proving certain inductive
properties of  $\varphi$  as we increase $\tau$  in discrete units.
First we introduce the deviation of  $\varphi$  from  $f_b$. It is
convenient to write this in the form
\be
\varphi (\xi,\tau) =  f^*_b(\xi)(1+e_b(\xi)\psi(\xi,\tau))
\en
where we introduced
\be
e_b(\xi)={1\over p-1 + b\xi^{2k}}.
\en
Then (1) is equivalent to
\be
\dot{\psi} = {\cal L}_{\tau}\psi + N(\psi) + D_\tau(\psi') +
P_{\tau}(\psi)
\en
where
\be
{\cal L}_{\tau} = L_{\tau}^{-2}\partial_{\xi}^2 - {1\over 2k}
\xi \partial_{\xi} + 1,
\en
the nonlinear term is given by
\be
N(\psi)= (1+e_b\psi)^p-1-pe_b\psi
\en
and
\BE
D_\tau(\psi')& =& -{4pkb\over p-1}L^{-2}_\tau e_b\xi^{2k-1}\psi'
\\
P_{\tau}(\psi) \
& =& L_{\tau}^{-2}\xi^{2k-2}e_b(\xi)(\alpha_1 + \alpha_2\xi^{2k}e_b
+(\alpha_3 + \alpha_4\xi^{2k}e_b)\psi) .
\EN
where $\alpha_i=\alpha_i(k,p,b)$ are constants. Note that, as opposed to
(3.41),
there is no potential in (7). This will greatly simplify the
analysis, see Lemma 1 below. Due to the factor $L_{\tau}^{-2}$, the
$ D_\tau$, $ P_\tau$ terms will be small,
like the "irrelevant" $F$ in Section 3.

Let
\be
M={2kp\over p-1}
\en
and
consider  $\psi (\sigma)$  of the form
\be
\psi (\sigma) = \sum\limits_{m=0}^{[M]}\psi_m(\sigma)h_m(\cdot,\sigma)
+ \psi^{\perp}(\sigma)\equiv \psi^< +\psi^\perp
\en
where $[\cdot]$ denotes the integer part and
$\psi^{\perp}$  is  $C^0$  in  $\xi$, is bounded by
\be
|\psi^{\perp}(\sigma)|_\sigma \equiv \sup_\xi
(L_{\sigma}^{-M} + |\xi|^{M})^{-1}
|\psi^{\perp}(\xi,\sigma)| \leq \varepsilon (\sigma)\en
and is orthogonal in  $L^2({\Bbb R},d\mu_{\sigma})$  to  $h_m(\sigma)$
with  $m \leq [M]$, where the scalar product is defined by (2.32).  Also
assume
\be
|\psi_m(\sigma)| \leq \left\{\begin{array}{ll}
\epsilon(\sigma) & m \neq 2k\\
\epsilon (\sigma)^{3/2} & m = 2k
\end{array}\right.
\en
and take
$$\varepsilon (\sigma) = L_{\sigma}^{-\delta}$$
for  $\delta > 0$. The reason for choosing $M$ larger than $2k$ is that
the integration by parts works only for such large $M$, see Lemma 1 below.
However, to control the nonlinear term (8), we cannot take $M$ too large.
We will need $N(\psi)$ to be bounded by $|\xi|^M$. But, with the choice
(11),
$(e_b\psi)^p\leq C|\xi|^{(M-2k)p} = C|\xi|^M$.
 We have then the

\vs{3mm}

\no {\bf Proposition.}  {\it Given  $\rho > 0$, there exist  $\delta > 0$
and
${\bar\tau} < \infty$  such that\ if  $\sigma > {\bar\tau}$  and  $\psi
(\sigma)$  satisfies} (12)--(14) {\it then the equation} (6)
{\it has a unique
classical solution, for  $\tau \in [\sigma,\sigma + \rho  ]$  which can be
expressed in the form} (12){\it with}
\be
|\psi_m(\tau) - e^{(\tau - \sigma)(1-{m\over 2k})}\psi_m(\sigma)|
\leq (\tau - \sigma)C(\rho)\varepsilon (\sigma)^2
\en
\be
|\psi^{\perp}(\tau)|_{\tau} \leq Ce^{-{1\over p-1}(\tau - \sigma)}
\varepsilon (\sigma).
\en

\vs{3mm}
Here and below, $C(\rho)\leq Ce^{c\rho}$.
We will now prove the Theorem, given the Proposition.

\vs{3mm}

\no {\bf Proof of Theorem 4.}  Let  $\varphi (\xi,\tau_0)$  be given by
(2).
We put (see (4))
\be
\psi (\xi,\tau_0) = \sum\limits_{i=0}^{2k-1} d_i\xi^i +
(p-1 + b\xi^{2k})^{p\over p-1}g(\xi) = \sum\limits_{m=0}^{[M]}
\psi_m(\tau_0)h_m(\xi,\tau_0) + \psi^{\perp}(\xi,\tau_0)\en
and get from (2.30)--(2.32),
\be
|\psi_m(\tau_0) - d_m| \leq C(L_{\tau_0}^{-1} +
L^m_{\tau_0}\varepsilon)\quad m < 2k ,
\en
\be
|\psi_m| \leq CL^m_{\tau_0}\varepsilon\quad m \geq 2k
\en
and, since
\be
\psi^{\perp} = \gamma - \sum\limits_{m=0}^{[M]} h_m(\gamma,h_m)_{\tau_0}
(h_m,h_m)^{-1}_{\tau_0}\en
with  $\gamma = (p-1 + b\xi^{2k})^{p\over p-1}g$, we find
\BE
|\psi^{\perp}(\xi,\tau_0)|& \leq& C\varepsilon [1 + |\xi|^{M}
+ \sum\limits_{m=0}^{[M]} (L_{\tau_0}^{-m} + |\xi|^m)L^m_{\tau_0}]
\non\\
&\leq& C\varepsilon L_{\tau_0}^{M}(L_{\tau_0}^{-M} +
|\xi|^{M})
\EN
where we use, see (2.30),
\be
|h_m(\xi,\tau)| \leq C(L_{\tau}^{-m} + |\xi|^m).
\en
Therefore, given  $\tau_0$,  we may, by taking  $\varepsilon$  small,
satisfy (12)--(14) for  $\sigma = \tau_0$.  Moreover, as in Section 3,
instead
of varying  $d_m$
in (1.14), we can vary  $\psi_m(\tau_0),\ m < 2k$, in (17),
in the interval defined by (14) with $\sigma=\tau_0$.
\medskip
Put now  $\tau_n = \tau_0 + n\rho$, and write
\be
\varphi (\xi,\tau_n) =  f_{b_n}(\xi)(1+e_{b_n}(\xi)\psi(\xi,\tau_n))
\en
with  $\psi$  satisfying (12)--(14) with  $\sigma = \tau_n$.  By the
Proposition,
$$\varphi (\xi,\tau_{n+1}) =
f_{b_n}(\xi)(1+e_{b_n}(\xi)\tilde{\psi}(\xi,\tau_{n+1}))
$$
with  $\tilde {\psi}$
satisfying
\be
|\tilde{\psi}_m - e^{\rho (1 - m/2k)}\psi_m| \leq C(\rho)\rho
\varepsilon (\tau_n)^2
\en
and
\be
|\tilde {\psi}^{\perp}|_{\tau_{n+1}} \leq Ce^{-{\rho\over p-1}}
\varepsilon (\tau_n).
\en
Put now
\be
b_{n+1} = b_n - \tilde {\psi}_{2k}\en
whereby, for $\delta$ small enough in the definition of $\epsilon(\sigma)$,
\be
\varphi (\xi,\tau_{n+1}) = f_{b_{n+1}}(\xi)(1+e_{b_{n+1}}(\xi)
\overline{\psi}(\xi,\tau))
\en
and
\BE
&&|\overline {\psi}_m - e^{\rho (1-m/2k)}\psi_m| \leq
C(\rho)\rho \varepsilon (\tau_n)^2\quad m < 2k\\
&&|\overline {\psi}_{2k}| \leq C \rho \varepsilon (\tau_n)^2 <
\varepsilon (\tau_{n+1})^{3\over 2}\\
&&|\overline {\psi}_m| \leq \varepsilon (\tau_{n+1})\quad m > 2k
\\
&&|\overline {\psi}^{\perp}|_{\tau_{n+1}} \leq \varepsilon (\tau_{n+1}).
\EN
We use $C(\rho) \varepsilon (\tau_n)^2 \leq \varepsilon (\tau_{n+1})$
and $C(\rho)\varepsilon (\tau_{n})^{3\over 2}\leq \varepsilon
(\tau_{n+1})$.
Moreover, by the same topological argument as in Section 3, we
now establish the
existence of  $\psi_m(\tau_0)$
for $m<2k$, such that  $\psi$  in (23) satisfies
(12)--(14) for all  $n$. The contrary assumption would now
allow us to construct a homotopy between the identity
and constant maps respectively from $S^{2k}\rightarrow S^{2k}$.
 From (26, 24, 14) we then deduce that
$$b_n \mathrel{\mathop{\longrightarrow}\limits_{n \to \infty}} b^*$$
and $|b^*-b_0|\leq C\varepsilon(\tau_0)^{3\over 2}$,
with $b_0=b$, implies that
$b^* \to b$  as  $\varepsilon \to 0$  and  $\tau_0 \to \infty$ (see (18,
19, 21)).
\hfill $\Box$

\vs{3mm}

We now turn to the proof of the Proposition. We consider the
following integral
equation related to our PDE (6):
\be
\psi (\tau) = K_{\tau \sigma}\psi (\sigma) + \int\limits_{\sigma}^{\tau}
dsK_{\tau s}(N(\psi (s)) + P_s(\psi(s)))+{\cal D}(\psi,\tau)
\en
where
\be
{\cal D}(\psi,\tau)=4pkb(p-1)^{-1}\int_\sigma^\tau ds L_s^{-2}
\int d\xi'\p_{\xi'}(K_{\tau s}(\xi,\xi')e_b(\xi'){\xi'}^{2k-1})
\psi(\xi', s)
\en
A classical solution of (6) satisfies (32). Note that ${\cal
D}(\psi,\tau)$
is obtained by integration by parts from the term that
naturally follows from the integration of (6). This form
is more convenient for us since we want to work with
$C^0$ data. We show that (32) has a unique solution in
a suitable space and that this solution is the classical
solution of (6).

Let us introduce the following norms in  $C^0({\Bbb R})$.
We write  $\psi\in C^0({\Bbb R})$  as in (12)
\be
\psi (\xi) = \sum\limits_{m=0}^{[M]} \psi_mh_m(\xi,\sigma) +
\psi^{\perp}(\xi)\en
and let
\be
\Vert \psi \Vert_{\sigma} = \sum_{m=0}^{[M]}
 |\psi_m| + |\psi^{\perp}|_{\sigma}.
\en
with $|\psi^{\perp}|_{\sigma}$ defined in (13).
It is straightforward to check that
\be
C_1(\sigma)\Vert \psi \Vert \leq \Vert \psi \Vert_{\sigma} \leq
C_2(\sigma)\Vert \psi \Vert\en
with  $C_i > 0$  and
\be
\Vert \psi \Vert = \sup\limits_{\xi}\ (1 + |\xi|^{M})^{-1}|\psi (\xi)|,
\en
so  $C^0({\Bbb R})$ is complete in the norm
$\Vert\cdot\Vert_\sigma$.

Write (32) as
\be
\psi (\tau) = \psi^0(\tau) + {\cal N}(\psi,\tau) +
{\cal P}(\psi,\tau)+{\cal D}(\psi,\tau)
\en
with (using (2.33)),
\be
\psi^0(\tau) = K_{\tau \sigma}\psi(\sigma) = \sum\limits_{m=0}^{[M]}
e^{(\tau - \sigma)(1 - {m\over 2k})}\psi_m h_m(\xi,\tau) + K_{\tau \sigma}
\psi^{\perp}.
\en
Let us first bound $\psi^0$:
\vs{3mm}

\no{\bf Lemma 1.}  {\it We have}
\be
\Vert K_{\tau \sigma}\psi^{\perp}\Vert_{\tau} \leq
Ce^{-{1\over p-1}(\tau - \sigma)}\Vert \psi^{\perp}\Vert_{\sigma}.
\en

\vs{3mm}

\no {\bf Proof} We use the conjugation (2.24): let
\be
\theta (\xi) = \psi^{\perp}(L_{\sigma}^{-1}\xi),\ \tilde {\theta}(\xi)
= (K_{\tau \sigma}\psi^{\perp})(L_{\tau}^{-1}\xi)\en
whence
\be
\tilde {\theta} = e^{(\tau - \sigma){\cal L}}\theta\en
and (see (35, 13)),
\be
|\theta (\xi)| \leq L_{\sigma}^{-M}(1 + |\xi|^{M})
\Vert \psi^{\perp}\Vert_{\sigma}\en
with
$$(\theta,h_m) = 0\;\;\; m \leq [M],$$
where the scalar product now is given by (2.10).
Proceeding as in the derivation of (3.67), we get, for  $\tau - \sigma \geq
1$,
\BE
|(e^{(\tau - \sigma){\cal L}}\theta)(\xi)| &=&
|\int \partial_{\xi'}^{[M]+1}N(\xi,\xi')f^{(-[M]-1)}
(\xi')d\xi'|\non\\
&\leq& Ce^{(\tau - \sigma)}e^{-([M]+1){\tau - \sigma\over 2}}
L_{\sigma}^{-M}(1 + |\xi|^{M})\Vert \psi^{\perp}\Vert_{\sigma}
\EN
where  $N(\xi,\xi')$  is defined in (3.63),
$f(\xi)=e^{-\xi^2/4}\theta(\xi)$
and we used the analogue of Lemma 3.6, i.e.
$$
|f^{(-m)}(\xi)|\leq Ce^{-m{\tau-\sigma\over 2}}L_\sigma^{-M}
(1+|\xi|)^{M-m}\Vert \psi^{\perp}\Vert_{\sigma}
$$
for $ 0\leq m \leq [M]+1$, and Lemma 3.4. Since by (11) and (2.4),
$$e^{-{M\over 2}(\tau - \sigma)}L_{\sigma}^{-M} = e^{-{p\over p-1}
(\tau - \sigma)}
L_{\tau}^{-M},
$$
we get
\be
|K_{\tau \sigma}\psi^{\perp}(\xi)| \leq Ce^{-{(\tau - \sigma)
\over p-1}}
(L_{\tau}^{-M} + |\xi|^{M}) \Vert \psi^{\perp}\Vert_{\sigma}.
\en
For  $\tau - \sigma < 1$  we need not integrate by parts and the bound
follows using (3.48).\hfill $\Box$

In particular, from Lemma 1 and (39, 13, 14) we deduce
$$\Vert \psi^0\Vert_{\tau} \leq C(\rho)\varepsilon (\sigma).$$
To solve (32), consider the ball
$$
{\cal B}=\{\psi\in C^0({\bf R}\times [\sigma,\sigma+\rho])\;\mid
\;\sup_{\tau\in[\sigma,\sigma+\rho]}\|\psi-\psi^0\|_\tau
\leq\epsilon(\sigma)^{\frac{1+{\tilde p}}{2}}\}
$$
where ${\tilde p}= \min (p,2)$.
Note that, for $\psi \in {\cal B}$, we have, for $s\in
[\sigma,\sigma+\rho]$
\BE
|\psi_m(s)|&\leq & C(\rho)\epsilon(\sigma)\non\\
|\psi^{\perp}(s)|&\leq & C(\rho)\epsilon(\sigma)
(L_{s}^{-M} + |\xi|^{M})
\EN
We have then

\vs{3mm}

\no {\bf Lemma 2.} {\it Equation} (32) {\it has a unique solution
$\psi\in{\cal B}$. The properties} (15){\it and} (16)
{\it hold for $\psi$.}

\vs{3mm}

\no{\bf Proof of Lemma 2.} We shall use the contraction mapping principle.
Let us first bound the nonlinear term
\BE
{\cal N}(\xi,\tau) &=& \int\limits_{\sigma}^{\tau} ds\int d\xi'
K_{\tau s}(\xi,\xi')N(\psi(s),\xi')\non\\
&=& \sum\limits_{m=0}^{[M]} {\cal N}_m(\tau)h_m(\xi,\tau)
+ {\cal N}^{\perp}(\xi,\tau).
\EN
Here (use (2.33))
\be
{\cal N}_m(\tau) = \int\limits_{\sigma}^{\tau} ds
e^{(\tau - s)(1 - {m\over 2k})}
{(h_m,N)_s\over (h_m,h_m)_s}.
\en
and
\be
{\cal N}^{\perp}(\cdot,\tau)=\int\limits_{\sigma}^{\tau} ds
K_{\tau s}N^\perp(\psi(s))
\en

\vs{3mm}

Consider first ${\cal N}_m(\tau)$.
Let $\chi(\xi)=\chi(|\xi|\leq 1)$ and $\chi^c=1-\chi$ and
insert
$$
N=\chi N+\chi^c N
$$
in (48). From (8) and the bounds (46) on $\psi$, we have
\be
|\chi^c N|\leq \chi^c(C(\rho)\epsilon(\sigma)
e_b|\xi|^{M})^{p}\leq  (C(\rho)\epsilon(\sigma))^p|\xi|^M
\en
(use (5) and recall that $M=2kp(p-1)^{-1}$) and therefore from (2.32)
\be
|{(h_m,\chi^cN)_s\over (h_m,h_m)_s}|
\leq e^{-L_\tau^2/c} \leq \epsilon(\sigma)^2 .
\en

As for $\chi N$, Taylor expanding, we obtain
\be
\chi N=\sum_{j=2}^K c_j\chi(e_b\psi)^j +R_K
\en
with the bound for the remainder
\be
|R_K|\leq C\chi|e_b\psi|^{K+1}\leq  (C(\rho)\epsilon(\sigma))^{K+1} .
\en
Thus,
\be
|{(h_m,R_K)_s\over (h_m,h_m)_s}|\leq
 L_s^m (C(\rho)\epsilon(\sigma))^{K+1}
\leq\epsilon(\sigma)^2
\en
provided we take $K>K(\delta)$.

For the sum in (52) insert the
decomposition (12):
\be
\chi(e_b\psi)^j=\chi(e_b\psi^<)^j+S_j
\en
and, by the bound (46) for $\psi^\perp$, we get, for
$S=\sum_{j=2}^K c_j S_j$
\be
|S| \leq  C(\rho)\epsilon(\sigma)^2\chi(\xi)
(L^{-M}_\tau+|\xi|^M).
\en
Thus
\be
|{(h_m,S)_s\over (h_m,h_m)_s}|\leq C(\rho)L_s^{m-M}\epsilon(\sigma)^{2}
\leq C(\rho)\epsilon(\sigma)^2 .
\en

For the first term in (55), we insert
\be
e_b=(p-1)^{-1}(\sum_{l=0}^L(-\frac{b\xi^{2k}}{p-1})^l+
(-\frac{b\xi^{2k}}{p-1})^{L+1}e_b(\xi)),
\en
and the expansion (12) for $\psi^<$,
to get
\be
A\equiv \sum_{j=2}^Kc_j\chi(e_b\psi^<)^j=
\sum_{{\bf n},p}c_{{\bf n},p}\chi\xi^p\prod_{i=1}^{[M]}
\psi^{n_i}_ih_i^{n_i}+\epsilon(\sigma)^2\xi^{2k(L+1)}\chi Q
\equiv A_1+A_2
\en
where ${\bf n}=(n_1,\dots,n_{[M]})$, $\sum n_i\geq 2$
and $\chi Q$ is bounded.

For $A_1$, put $\chi=1-\chi^c$, estimate the $\chi^c$-term
by (51) and use
\be
|{(h_m,\xi^p\prod h_i^{n_i})_s\over (h_m,h_m)_s}|
=\left\{ \begin{array}{ll}
0 & p + \sum n_i < m\\
\leq CL_s^{m-p-\sum n_i} & p + \sum n_i \geq m
\end{array}\right.
\en
and (46), to get
\be
|{(h_m,A_1)_s\over (h_m,h_m)_s}|
\leq  C(\rho)\epsilon(\sigma)^2 .
\en
For $A_2$, take $2k(L+1)\geq M$ to obtain (61) again, for $A_2$.
Thus, inserting (51, 54, 57, 61) into (48),
\be
|{\cal N}_m(\tau)|\leq (\tau - \sigma)C(\rho)\epsilon(\sigma)^2
\en
for $\tau\in [\sigma,\sigma+\rho]$.

\vs{3mm}

For ${\cal N}^\perp$ in (47, 49), we proceed in a similar fashion; (50)
yields
\be
|(\chi^c N)^\perp|\leq ( C(\rho)\epsilon(\sigma))^p
(L^{-M}_\tau+|\xi|^M)
\en
using $X^\perp=X-\sum_0^{[M]}h_m(h_m,h_m)_s^{-1}(h_m,X)_s$,
(3.74), and the first inequality in (51).

For $(\chi N)^\perp$, we write as in (52, 55, 59),
\be
\chi N= A+S+R_K .
\en
$S$ has the bound (56), and, from (53), we get, for $K$ large,
\be
|R_K|\leq \epsilon(\sigma)^2L_\sigma^{-M}.
\en
As for $A$, by (59), $A_2$ satisfies a bound like (56)
 and for $A_1$
we divide the sum into $p +\sum n_i > [M]$ and $\leq[M]$. For the first
sum,
\be
\chi|\xi^p\prod_{i=1}^{[M]}
h_i^{n_i}|\leq C(L_s^{-M}+|\xi|^M)
\en
since $h_i$ is bounded by (22),
and, for the second sum, replace $\chi=1-\chi^c$ by $-\chi^c$ since
$1$ will not contribute to $A^\perp$ (
$(\xi^p\prod_{i=1}^{[M]}
h_i^{n_i})^\perp =0$ for these terms). Since $|\xi|\geq 1$,
in this case
\be
\chi^c|\xi^p\prod_{i=1}^{[M]}
h_i^{n_i}|\leq C|\xi|^M
\en
and altogether
\be
|A^\perp|\leq  C(\rho)\epsilon(\sigma)^2(L_s^{-M}+|\xi|^M)
\en
(we used again the fact that this bound for $A$ implies it for $A^\perp$).
By (63),(65),(68),
\be
|N^\perp|\leq  C(\rho)(\epsilon(\sigma)^2+\epsilon(\sigma)^p)
(L_s^{-M}+|\xi|^M)
\en
and therefore, using (41, 42) and Lemma 3.4,
\be
|{\cal N}^\perp|\leq  C(\rho)(\tau -\sigma)
(\epsilon(\sigma)^2+\epsilon(\sigma)^p)
(L_{\sigma}^{-M}+|\xi|^M).
\en
We may bound $ C(\rho)(\tau -\sigma)
(\epsilon(\sigma)^2+\epsilon(\sigma)^p)$ by
$\epsilon(\sigma)^{\frac{{\tilde p}+1}{2}}$.

\vs{3mm}

Next, we consider $\cal P$ in (38). From (10) we have
\be
|P_\tau\chi^c|\leq CL_\tau^{-2}|\xi|^{M-2}
\en
which fits to our bounds (see (63)). The $|\xi|<1$ analysis proceeds as
above.

\vs{3mm}

Finally, for the $\cal D$ term in (33), the terms where $\p_{\xi'}$
does not act on  $K$ are estimated as above and
we write the remaining
term as $4pkb(p-1)^{-1}F$ with
\be
F(\xi)=
\int\limits_{\sigma}^{\tau} dsL_s^{-2}\int d\xi'
e^{(\tau - s)\over 2k}\partial_{\xi}
K_{\tau s}(\xi,\xi')\rho(\xi',s)=\sum_{m=0}^{[M]} F_m h_m +F^\perp
\en
where $\rho(\xi',s)=
(\xi')^{2k-1}e_b(\xi')\psi
(\xi',s)$,
and where we used (2.27, 2.28) to change $\partial_{\xi'}$ into
$\partial_{\xi}$. We write again
\be
\rho=\sum_{m=0}^{[M]}\rho_m h_m+\rho^\perp
\en
and we have from (46),
\be
\|\rho(s)\|_s\leq C(\rho)\epsilon(\sigma).
\en
Since $\p_\xi h_m=mh_{m-1}$ and the adjoint of $\p_\xi$ in
$(\cdot ,\cdot)_\tau$ is $-\p_\xi +{L_\tau\over 2}\xi$, we have
\be
|F_m(\tau)|\leq (m+1)\int_\sigma^\tau ds L_s^{-2}
e^{(1-{m\over 2k})(\tau-s)}
|\rho_{m+1}(s)|+(\tau - \sigma) C(\rho) L_{\sigma}^{-1} (h_m,
\xi K_{\tau s}\rho^\perp)_\tau (h_m,
h_m)_\tau^{-1} .
\en
Therefore
\be
|F_m(\tau)|\leq (\tau-\sigma)C(\rho)L^{-1}_\sigma\epsilon(\sigma)
\leq (\tau-\sigma)\epsilon(\sigma)^2 .
\en
As for $F^\perp$, we have
\be
F^\perp = \int_\sigma^\tau ds L_s^{-2}e^{(\tau - s)\over 2k}
\p K_{\tau s}\rho(s)^\perp
\en
and we need the bound for $\p K$, derived from (2.27, 2.28):
\be
|\partial_{\xi}K_{\tau s}(\xi,\xi')| \leq
C(\rho)L\delta_{{1\over 2}L^2}(e^{s-\tau\over 2k}\xi-\xi')
\leq{C(\rho)L_{\tau}\over \sqrt {\tau - s}}
\delta_{{1\over 2}L^2}(e^{s-\tau\over 2k}\xi-\xi')
\en
(where, recall, $L^2=L^2_s(1-e^{s-\tau})^{-1})$. This implies
\be
|F^\perp|_\tau\leq C(\rho)L_\sigma^{-1}\epsilon(\sigma)\leq
\epsilon(\sigma)^2
\en
The bounds (62, 70), (76, 79), and similar
bounds for the other terms in $\cal P$, $\cal D$,
 imply that the RHS of (32) maps
$\cal B$ into itself. That this map also contracts
in this ball is showed in a similar fashion. Finally, (15)
follows from (32, 39, 62, 76) and (16) follows from
(32, 40, 70, 79), using
$p>1$ in (70) to bound its RHS by
$e^{-\frac{\rho}{p-1}}\epsilon(\sigma)(L^{-N}_\tau+ |\xi|^N)$.
 \hfill $\Box$

\vs{3mm}

To conclude the proof of the Proposition, we need to show
that the $\psi$ as constructed above actually is $C^2$
in $\xi$ and $C^1$ in $\tau$ on $(\sigma,\sigma+\rho]$.

\vs{3mm}

For smoothness, the idea is to improve the smoothness of
$\psi$ iteratively using (32) and the regularity properties
of the kernel $K$. This is completely straightforward, but
we will sketch the argument here for completeness.
Let us consider the least regular term
in (32) i.e. the $F$ term in (72).
The bound (78) may be improved to
\be
|\partial_{\xi}K_{\tau s}(\xi_1,\xi')
-\partial_{\xi}K_{\tau s}(\xi_2,\xi')| \leq
{C(\rho)L_{\tau}^{2-\eta}|\xi_1-\xi_2|^{1-\eta}\over
(\tau - s)^{1-{\eta\over 2}}}
(\delta_{{1\over 2}L^2}(e^{s-\tau\over 2k}\xi_1-\xi')+
\delta_{{1\over 2}L^2}(e^{s-\tau\over 2k}\xi_2-\xi'))
\en
for $\eta>0$.
This implies that $F$ is $C^{1-\eta}$ (H\"older continuous
with exponent $1-\eta$).
The other terms of $\cal D$, $\cal N$, and $\cal P$
are analyzed similarily to be $C^{2-\eta}$.
Thus our solution $\psi\in C^{1-\eta}$, and, by considering the solution
$\psi^0$ of the linear equation, we see that its ${1-\eta}$
H\"older norm is bounded
by ${\cal O}((s-\sigma)^{-{1-\eta\over 2}})$ for $s-\sigma$ small.
 With this knowledge
we next prove that $F\in C^{2-\eta}$. Indeed,
\be
F(\xi)=
\int\limits_{\sigma}^{\tau} dsL_s^{-2}\int d\xi'H(\xi,\xi')
\en
where, putting the derivative back on $\xi'$, and using
$\int d\xi'\partial_{\xi'}
K_{\tau s}(\xi,\xi')=0$, we have
\be
H(\xi,\xi')=
\partial_{\xi'}
K_{\tau s}(\xi,\xi')(\rho(\xi',s)-\rho(e^{\tau-s\over
2k}\xi,s))
\en
and we may now use the H\"older property of $\rho$
(for each power of $|\xi'-e^{\tau-s\over
2k}\xi|$, one gains a power of $|\tau -s|^{1/2}$) to bound
\be
\int d\xi'|\p_\xi H(\xi,\xi')|\leq C(\rho)(\tau-s)^{-{1+\eta\over 2}}
(s-\sigma)^{-{1-\eta\over 2}}
\en
where the $(s-\sigma)^{-{1-\eta\over 2}}$ comes from the H\"older
estimate on $\psi$. Now, (83) is integrable in $s$, for $\eta$ small.
Thus $\psi$ is $C^1$. This allows us
to integrate by parts and to write (33) as
:
\be
{\cal D}(\psi,\tau)=-4pkb(p-1)^{-1}\int_\sigma^\tau ds L_s^{-2}
\int d\xi'K_{\tau s}(\xi,\xi')e_b(\xi'){\xi'}^{2k-1}
\p_{\xi'}\psi(\xi',s)
\en
and by the above argument to prove that $\cal D$
and hence $\psi$ is $C^{2-\eta}$.

Finally, we write (38), reasoning as in (82),
\be
\p\psi(\xi,\tau)=\p\psi^0(\xi,\tau) +\int ds\int d\xi' \p_\xi
K_{\tau s}(\xi,\xi')(\Psi(\xi')-
\Psi(\xi))
\en
with $\Psi \in C^{1-\eta}$ and apply the argument above to conclude
that $\psi$ is $C^2$. The derivative with respect to $\tau$ is similar.
This concludes the proof of the
Proposition.\hfill $\Box$

\end{document}